# Cryo scanning transmission X-ray microscope optimized for spectrotomography

A. F. G. Leontowich,[1] R. Berg,[1] C. N. Regier,[1] D. M. Taylor,[1] J. Wang,[1] D. Beauregard,[1] J. Geilhufe,[1] J. Swirsky,[1] J. Wu,[2] C. Karunakaran,[1] A. P. Hitchcock,[2] and S. G. Urquhart[3]

1. Canadian Light Source Inc., Saskatoon, Saskatchewan S7N 2V3, Canada

2. Department of Chemistry & Chemical Biology, McMaster University, Hamilton, Ontario L8S 4M1, Canada

3. Department of Chemistry, University of Saskatchewan, Saskatoon, Saskatchewan S7N 5C9, Canada

**Abstract:**

A cryo scanning transmission X-ray microscope, the cryo-STXM, has been designed and commissioned at the Canadian Light Source synchrotron. The instrument is designed to operate from 100 – 4000 eV ($\lambda$ = 12.4 – 0.31 nm). Users can insert a previously frozen sample, through a load lock, and rotate it ±70° in the beam to collect tomographic data sets. The sample can be maintained for extended periods at 92 K primarily to suppress radiation damage, and a pressure on the order of $10^{-9}$ Torr to suppress sample contamination. The achieved spatial resolution (30 nm) and spectral resolution (0.1 eV) are similar to other current soft X-ray STXMs, as demonstrated by measurements on known samples and test patterns. The data acquisition efficiency is significantly more favorable for both imaging and tomography. 2D images, 3D tomograms and 4D chemical maps of automotive hydrogen fuel cell thin sections are presented to demonstrate current performance and new capabilities, namely cryo-spectrotomography in the soft X-ray region.

Corresponding author: adam.leontowich@lightsource.ca

## 1. Introduction:

Scanning transmission X-ray microscopy (STXM) is an established imaging technique for investigating the structure and composition of matter at the micro- to nanometer length scale. Briefly, in STXM an X-ray beam is focused down to a small spot onto a sample using optics such as a zone plate (ZP) lens or Kirkpatrick-Baez (KB) mirrors [1-3]. Two dimensional images are compiled from a series of line scans, performed by scanning the sample downstream of the optic(s), or by scanning the optic(s) upstream of the sample. Modern soft X-ray STXMs, such as the ambient-STXM at the Canadian Light Source (CLS, Saskatoon, Canada) spectromicroscopy (SM) beamline 10ID-1 [4], acquire images of sub-micrometer thick specimens with a bright field spatial resolution that is close to diffraction limited [5]: 30 nm half pitch is now considered routine at most facilities. When a STXM is combined with a soft X-ray beamline at a synchrotron, spectromicroscopy is enabled [2]. The focused beam can be held at one spatial location while the photon energy is scanned to perform X-ray absorption near edge spectroscopy (XANES) on a small volume [2]. More often, a series of monochromatic images over a photon energy range of interest is collected, and each pixel in the resulting "image stack" contains a XANES spectrum [6]. Image stacks can be converted into quantitative chemical maps using thickness normalized reference spectra [2]. Soft X-ray STXMs have found applications in diverse fields such as human health [7], agriculture [8], environmental studies [9], nanofabrication [10], and advanced materials [11].

The overall form and function of current ZP based STXMs can be traced back to instruments built in the 1980s at the National Synchrotron Light Source (NSLS, Brookhaven National Laboratory, USA) [12,13]. A concise timeline of instrumentation developments from around the world has been covered by Kilcoyne [14]. Several founders of the technique predicted that the combination of high spatial and photon energy resolution in the soft X-ray region, along with the scanning configuration where all inefficient optics are upstream of the sample, would have a natural advantage for low dose (J kg$^{-1}$) interrogation of organic matter, particularly biological specimens in the natural hydrated state [15]. This low-dose-low-damage potential was realized by the 1990s [16], and the STXM user community and number of operational instruments flourished soon after. The status of present instruments and several emerging applications have been recently reviewed by Hitchcock [3,17].

STXM instrumentation and software has become increasingly complex as researchers seek to extract ever more information from their samples, by exposing them to a variety of non-ambient conditions in-situ and/or using a variety of detection modes. For example, STXM tomography was first demonstrated in 1994 [18]. Imaging the same spatial region multiple times over a series of tilt angles typically results in a higher dose to the sample relative to a single transmission image [19]. Spectro-tomography, or multiple tomographs collected at multiple photon energies [20-22], results in an even higher dose. Other high dose modes which are becoming routine include electron yield [23,24], X-ray fluorescence [25], and ptychography [26]. Improving the spatial resolution is a constant research theme [27]. However, there is a proven link between spatial resolution and dose [19,28]. High dose often induces high radiation damage [28], especially for organic matter [16,29], and this leads to artefacts which can invalidate the measurement. The CLS SM user community wanted to perform and expand on these and other high dose measurements, especially tomography of hydrated organic/inorganic composites [22,30], without the high level of radiation damage artefacts typically associated with them.

Cooling samples to near liquid nitrogen (LN$_2$) temperatures of around 100 K or "cryo-cooling" is an established method to temporarily suppress some aspects of radiation damage [28,31,32], even in the field of X-ray microscopy [33,34]. Several cryo full field X-ray microscopes (TXM) are operational [35-37], and one soft X-ray cryo-STXM operated at NSLS from 1996 to 2011 [34,38,39]. We decided to pursue a cryo-cooled sample STXM as the best available option for a high-dose-low-damage soft X-ray microscope with excellent spectroscopic capabilities. In addition to the benefits for radiation sensitive samples, near LN$_2$ temperatures can suppress artefacts associated with sublimation of high vapour pressure substances under vacuum [40]. Cryo-cooled samples often give higher diffraction intensities especially at high angles via a reduction in Debye-Waller factors [41], and some intriguing chemical and physical phenomenon such as magnetic phase transitions and superconductivity are only accessible at near LN$_2$ temperatures [24].

While retrofitting a rotation stage into an existing STXM to perform tomography can be relatively simple [18,21,22], retrofitting a device to cryo-cool the sample can require extensive modifications. The condensation of water vapor and contaminants onto a cold sample [24,42] must be suppressed, typically by flowing cold dry gas over the sample [31,37], or placing the sample and other key components inside a vacuum chamber at a pressure $\leq 10^{-7}$ Torr. Instead of retrofitting, we began the design of an all-new instrument, drawing on experience gained from operating a STXM at the CLS since 2007. The major goals were to realize a STXM optimized for cryo-spectrotomography that could operate from 100 – 4000 eV, fully covering the soft X-ray region while touching the extreme ultraviolet and also the tender/hard X-ray regions.

Specifically, the instrument was required to insert a frozen sample into a vacuum chamber through a load lock, and then rotate it ±70° in the beam for tomography, all at near LN$_2$ temperature. In addition, it had to be able to perform the majority of tasks that are expected of modern soft X-ray STXMs.

## 2. Design

### 2.1 Initial boundary conditions

The ability to insert a frozen sample through a load lock, and rotate it ±70° in the beam, all at near LN$_2$ temperature is technically quite challenging. We selected a goniometer from a JEOL JEM-2010F transmission electron microscope (TEM), combined with a Gatan 630-DH high tilt tomography cryotransfer system [43,44] to provide these essential capabilities. This approach of adapting proven TEM technology to X-ray microscopy is similar to that used in the NSLS cryo-STXM [34,38,39]. While the NSLS cryo-STXM acquired images by scanning the TEM sample holder through a stationary focused beam, our approach to acquire images is by scanning the ZP/focused beam upstream of a stationary TEM sample holder. Coarse sample positioning is also possible by manipulating the TEM sample holder using the out of vacuum stepper motors of the goniometer. The Gatan 630 sample holder is advertised as having a "minimum temperature attainable in the [electron] microscope" of ≤103 K [44], which is suitable for suppressing radiation damage.

The new cryo-STXM is located downstream of the ambient-STXM, on the existing STXM branch of the SM beamline [4]. The downstream location provides an increased spatially coherent area available for ZP scanning for a given exit slit width and photon energy. This location also allowed installation and commissioning work to occur with minimal interruption to the SM user program. The ambient-STXM will continue with its current program, which includes hydrated samples as well as materials, prototype experimental setups and detectors which are not compatible with the vacuum system of the cryo-STXM. A photograph of both instruments inside the SM STXM hutch is presented in **Figure 1**.

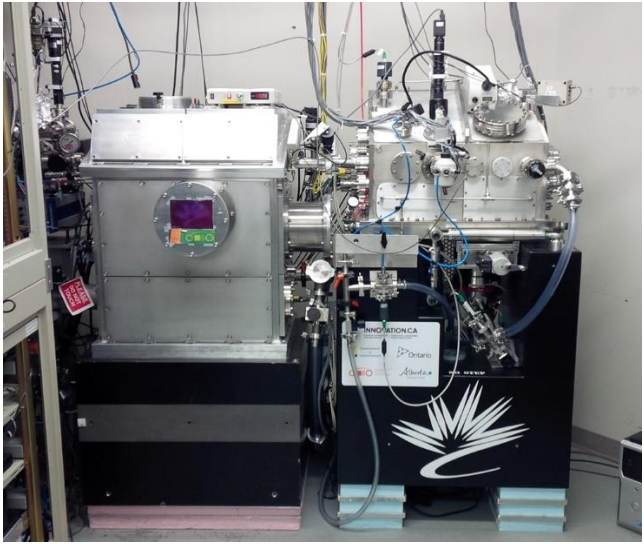

**Figure 1**: Photograph of the inboard side of the soft X-ray STXMs at the CLS SM beamline, ambient-STXM (left, upstream) and cryo-STXM (right, downstream).

## 2.2 Components on the optical axis

A soft X-ray ZP based STXM requires four main components on the optical axis: A ZP with integrated central stop, an order sorting aperture (OSA), a sample, and a transmission detector. These components are depicted in **Figure 2**.

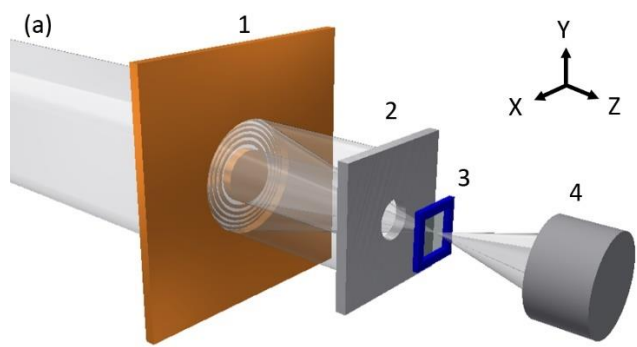
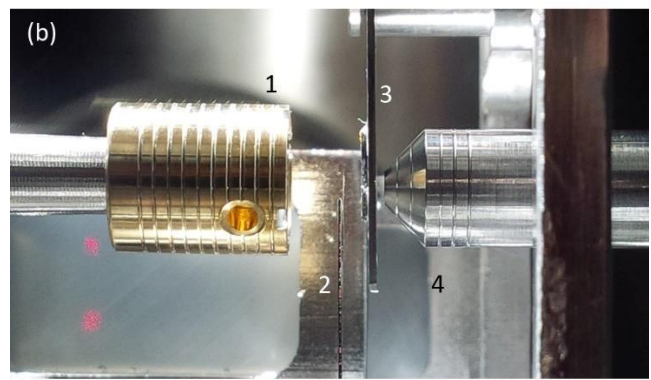
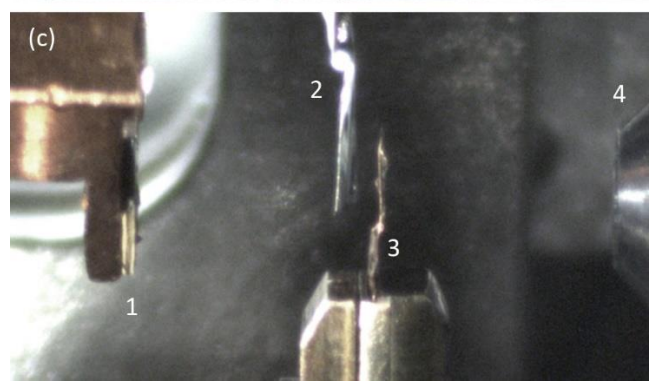

**Figure 2**: (a) Schematic depicting the four main components of a STXM: 1) ZP, 2) OSA, 3) sample, 4) transmission detector. (b) Photograph of the main components in conventional mode (inboard side view, through a viewport), with positions set for operation at 690 eV. The scribed lines on the ZP holder and detector cover have a pitch of 1.0 mm. (c) Photograph of the main components in goniometer mode (top view, using a microscope through a viewport), with positions set for operation at 690 eV. Beam enters from left in all sub-panels.

The soft X-ray source for the cryo-STXM is a rectangular aperture defined by horizontal and vertical slits at the end of the STXM branch of the SM beamline (a secondary source). The distance between the ZP and the slits is 4.3 m. A 0.75 mm × 0.75 mm × 100 nm silicon nitride window 3.0 m downstream of the slits isolates the vacuum system of the ambient-STXM from the beamline, and also limits the maximum illuminated area available for zone plate scanning in the cryo-STXM. We observed that this area is uniformly illuminated by the secondary source using a phosphor screen and camera inside the ambient-STXM. Further details of the beamline and performance have been reported elsewhere [4].

ZPs are purchased from outside sources. While the diameters specified typically fall in the range of a few hundred micrometers, the structure itself is supported on a silicon nitride membrane, framed by a silicon chip [5]. Current ZP chip dimensions are typically 6.5 mm × 6.5 mm, 5.0 mm × 5.0 mm, or 3.0 mm × 3.0 mm. These dimensions become problematic for achieving ZP to sample distances <3.0 mm with the Gatan 630 cryo sample holder. The design goal was to operate over the photon energy range 100 – 4000 eV while maintaining 30 nm half pitch spatial resolution with a ZP diameter ($D$) ≤300 µm. At 100 eV, the focal length of a 240 µm $D$, 25 nm outermost zone width ($\Delta r$) ZP with 90 µm diameter integrated central stop, used throughout this report, is approximately 0.5 mm [1]. A chip with standard dimensions would collide with part of the Gatan 630 well before reaching this focal length. We designed a new chip to enable operation at focal lengths <3.0 mm, which was then fabricated by Applied Nanotools. This chip begins as 3.0 mm × 3.0 mm, and then one edge is trimmed back by ≥0.7 mm. These chip dimensions are also advantageous for tilted samples and fluorescence detection. One ZP can be mounted at a time.

The cryo-STXM has two independent sample environments. One is JEOL 2010 pattern TEM sample holders, such as the Gatan 630, which are supported by the goniometer. We refer to this as "goniometer mode". The sample grids used are a custom design IFR-1 optimized for tomography [35], and one grid can be mounted at a time. The second sample environment is trapezoidal aluminum plates, used in the ambient-STXM and many other STXMs worldwide [14]. Up to six standard 3.0 mm diameter TEM grids, or six silicon nitride windows can be affixed to a single plate and positioned at the X-ray focus. The plate is supported by a three pin kinematic mount on top of a stage stack (§2.3). We refer to this as "conventional mode".

The OSA is a laser drilled hole in a 25 µm thick 304L stainless steel foil fabricated by Ladd Research Industries, positioned between the ZP and sample [13,14]. The foil is oriented horizontally for goniometer mode (Fig. 2c) [34], and vertically for conventional mode (Fig. 2b). A single foil can have several apertures of varying diameters available for in-situ selection. The foil is electrically isolated to enable biasing or electron detection via drain current. Sometimes the OSA is not required if the sample has appropriate dimensions [18]. The bright-field transmission detector is a CLS built scintillator-photomultiplier tube (PMT) device that provides a saturation count rate on the order of 40 MHz. Further details of this detector have been published elsewhere [45].

**2.3 Motion**

The components on the optical axis have motion requirements which are satisfied by commercially sourced stages located in and out of vacuum. Specifics of the stages are summarized in **Supplemental Table I**, while the stage layout is depicted in **Figure 3.**

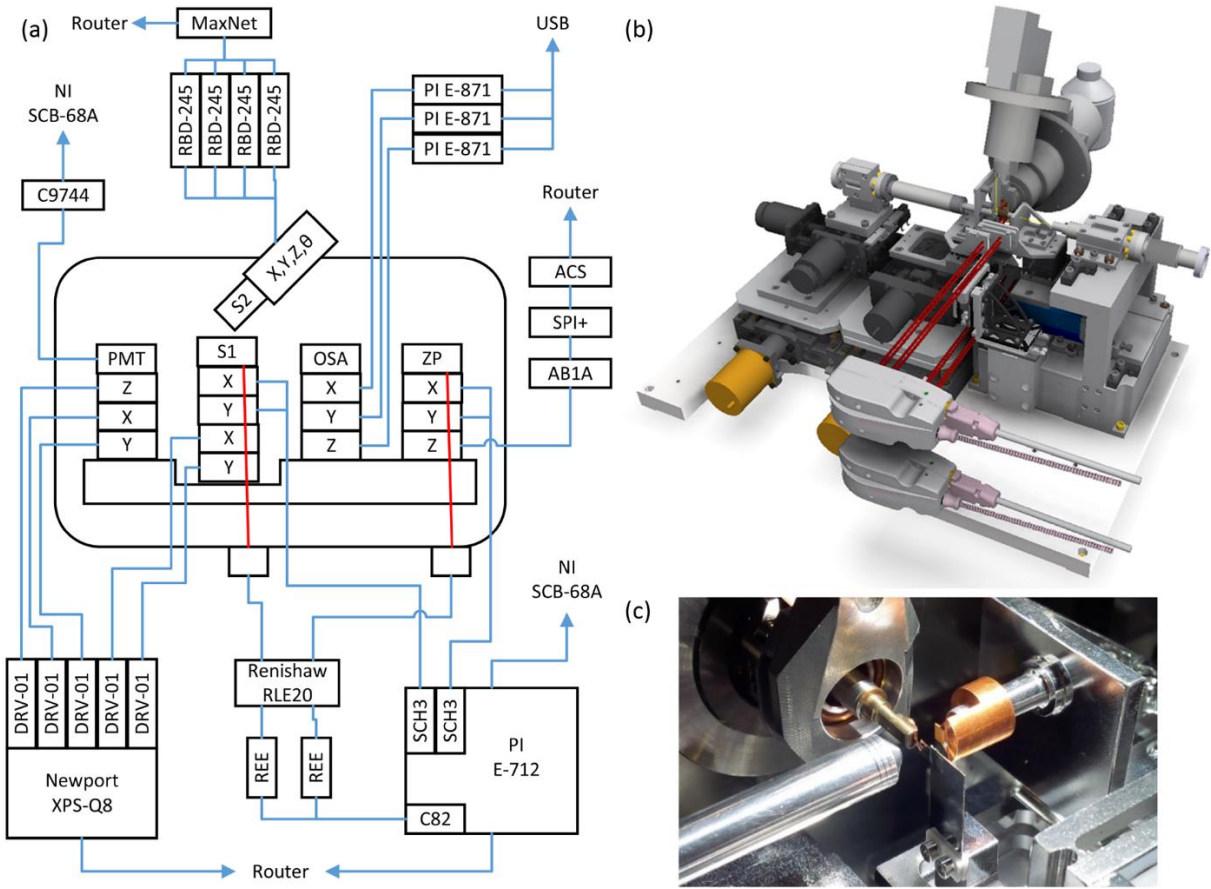

**Figure 3**: (a) Block diagram depicting the stage layout along with drivers and controllers. (b) 3D CAD model of the instrument with the vacuum chamber removed. (c) Photograph inside the vacuum chamber (outboard side view) in goniometer mode. Beam enters from the right in all sub-panels.

The rationale behind the stage layout is largely out of necessity after committing to the JEOL goniometer as the solution for cryo-tomography. This multi-axis stage provides sample motion in X, Y, Z, and θ (roll about X axis), but the full travel ranges are limited to X: 2.2 mm, Y: 1.5 mm, and Z: 1.0 mm. While θ is technically unlimited, we limit it to a practical range of ±70°. The 1.0 mm travel range for sample Z has implications for ZP Z and OSA Z motions. To maintain a sharp focus on the sample over the desired photon energy range of 100 – 4000 eV, the ZP to sample distance must have a range of about 20 mm. Since the goniometer sample Z travel range is only 1.0 mm, and is reserved for eucentric height adjustment when performing tomography [46], it was therefore necessary for a ZP Z stage to supply all of the travel range required to enable focusing. The optimal Z position of the OSA also varies with photon energy. For a 20 mm ZP Z travel range, the OSA to sample travel range must be about 4 mm. Again, since the goniometer cannot supply 4 mm of sample Z range, the OSA required a Z stage to supply all of the travel range needed. The OSA also required XY motion to align an aperture transverse to the focused beam, to reposition the aperture during large area ZP scans (**Supplemental Figure 1**), and to enable in-situ selection of apertures. Since the ZP and OSA have generous Z motion ranges for focusing, it was not necessary to incorporate a very large Z stage underneath the conventional sample and detector stages [4,14].

The JEOL goniometer was designed for sample positioning; sample scanning with the nanometer accuracy and speed required for a modern STXM is not available. To enable accurate and fast scanning microscopy, we placed the ZP on two piezo flexure stages, and rely on ZP scanning to form sub-100 µm × 100 µm images in goniometer mode (90 µm × 90 µm demonstrated in **Supplemental Figure 1**). ZP scanning can also be performed in conventional mode. The MAXYMUS STXM at BESSY II has previously demonstrated ZP scanning at close to diffraction limited spatial resolution with fields of view up to 20 µm × 20 µm [47]. The goniometer is removed from the instrument during conventional mode operation, and the changeover between sample environments takes about 4 hours. Lastly, the detector required coarse XYZ motion to align it in the transmitted beam. A load capacity of 3 kg and a long X range was realized to support and select a future area detector.

**2.4 Vibration and stability**

In order to achieve half pitch spatial resolution of ≤30 nm it is necessary to suppress sources of drift and vibrations in the instrument, especially those involving motion of the ZP and the sample relative to each other. As in many previous STXMs, a differential laser interferometer position encoder system (RLE20, Renishaw) is employed as part of a feedback loop. A set of mirrors above the ZP piezo stages represents the ZP XY position. These mirrors have a functional Z length of 25 mm. Another set of mirrors above the sample piezo stages represents the conventional mode sample XY position. In goniometer mode, it is not practical to have mirrors with the area required on the millimeter sized tip of the TEM sample holder, so a third set of mirrors on an internal reference frame was designed to represent goniometer mode samples. However, these have been found to be unnecessary in practice. The conventional sample mirrors are currently used to close the feedback loop for goniometer mode, and we rely on the intrinsic stability of the goniometer and vacuum chamber. Except for the mirrors, all of the components and adjustments of the RLE20 system are conveniently out of vacuum.

The stainless steel vacuum chamber for the cryo-STXM consists of a reinforced box lid atop a 50 mm thick baseplate. The interferometer heads and the goniometer are mounted directly to the walls of the lid. Finite element analysis (FEA) calculations were performed for chamber deflection under vacuum, and also for vibration analysis. Those parameters were optimized and the results have been reported elsewhere [48]. The baseplate is attached to a 1200 kg polymer concrete block (Basetek) via six adjustable struts [49], and the block rests on four passive dampers composed of extruded Styrofoam® in a sandwich configuration [50,51]. The turbopump is a magnetic bearing water cooled variant (§2.6), and the rough vacuum pumps are located outside of the SM beamline STXM hutch, about 2 m away from the STXMs on a different section of the concrete floor [52] separated by an expansion joint.

For thermal stability, all of the stages in vacuum are supported on an internal 22 mm thick Invar 36 plinth. The temperature at the sample position of the Gatan 630 sample holder is monitored and adjusted by a Gatan SmartSet 900 controller [44]. The controller can maintain the sample temperature between 92 K and 363 K with an accuracy of 0.1 K. The hutch has only coarse temperature control and limited acoustic control. Visiting researchers and staff sit in the hutch within 1 – 2 m of the STXMs while operating, which usually has no effect on the quality of the data.

## 2.5 Control system

A block diagram of the control system is presented in **Figure 4**.

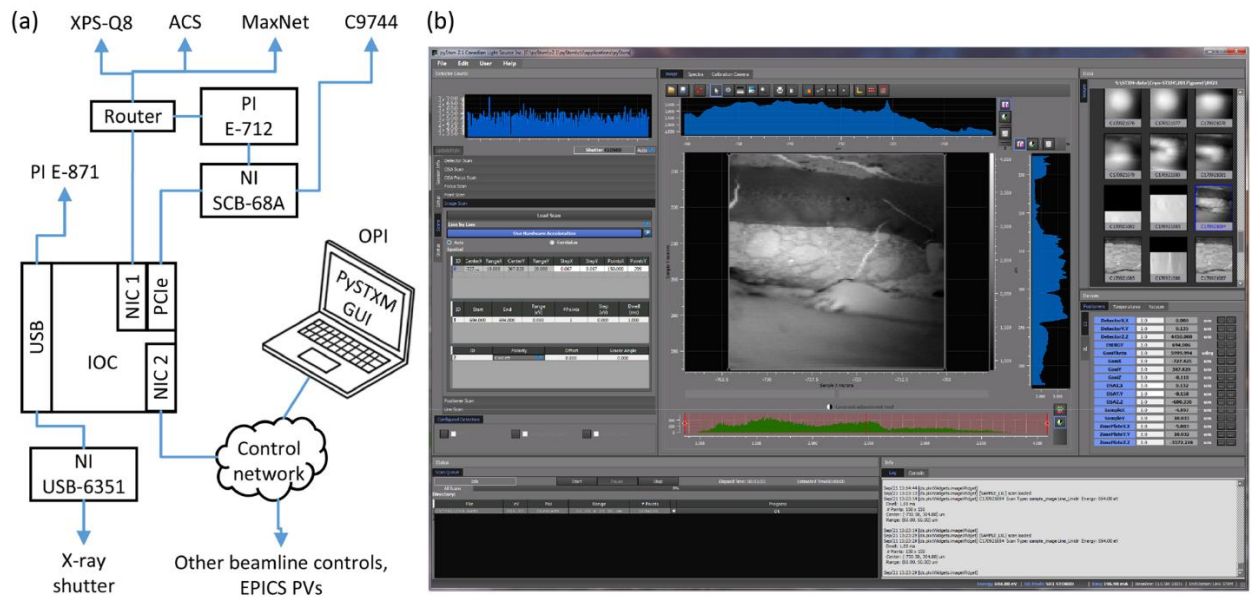

**Figure 4**: (a) Block diagram depicting the control system. (b) Screen capture of the graphical user interface of the control program, pySTXM.

The motor controllers and detector electronics are controlled by EPICS [53] (distributed control system) applications running on a Windows 7 64 bit input output controller (IOC), and are physically connected either by Ethernet, USB or as a National Instruments PCIe-6351 multifunction input/output card. All device control, feedback, and sequencing to collect data are implemented as EPICS drivers/applications, which are then accessed over the network via EPICS Channel Access. A data collection and graphical user interface program, pySTXM, was created in-house in Python and the Qt application framework. pySTXM runs on another operator interface (OPI) computer which accesses the cryo-STXM IOC as well as other devices along the beamline using the PyEpics module [54] over EPICS Channel Access.

pySTXM allows users to review the status of cryo-STXM and beamline parameters, calibrate and tune stages, align the components on the beam axis, initiate scans and acquire data, present the data to the user as it is being acquired, and to write data files. Users can load previously saved scan definition files, and review the estimated time required to execute a scan. Data is written to a network drive as a scan progresses, and users can pause a scan, or abort a scan with the option to save or discard the data collected to that point. Data files are written using the NXstxm definition of the NeXus scientific data format [55], and can be processed in the programs aXis2000 [56] and MANTiS [57] which are supported by SM beamline staff. Remote access is enabled via third party software.

Many types of scans have been implemented in pySTXM. Detector and OSA XY scans are available to align the detector and OSA on the beam axis, respectively. Focusing is accomplished by scanning the sample or ZP in XY while stepping ZP Z through a range. This is often aided by a calibrated out of vacuum optical microscope (Fig. 2c). Coordinated ZP-OSA scans are available for large area ZP scans (**Supplemental Figure 1**). The program automatically adjusts the ZP to sample distance to maintain focus during scans involving multiple photon energies. The undulator gap and phase, and the position and angles of beamline optics are also automatically updated to maintain high flux and photon energy resolution, or to perform automated polarization dependent measurements. The shutter automatically closes between scans to minimize the dose to the sample.

Scans involving the sample or ZP XY stages incorporate a feedback loop involving the RLE20 interferometer. The outputs of the RLE20 are two analog one volt peak-to-peak signals that represent the difference in position between the sample and ZP in X and Y (§2.4). These signals are interpolated and converted in hardware (REE0200A20B, Renishaw) to 0.79 nm resolution digital AquadB, and run directly into the PI E-712 piezo stage controller, closing a feedback loop in hardware between the ZP and sample XY positions with a servo update rate of 20 kHz. When activated, one component becomes the reference position, depending on whether the sample or ZP is going to be scanned. The E-712 determines how to drive the piezo stages supporting the other component, servoing them using the interferometer encoder feedback. Closing this feedback loop compensates for drift leading to image distortions, vibrations which degrade spatial resolution, and drift during scans covering multiple photon energies due to run out of the ZP Z stage [38]. The feedback/servo control is turned off before, and on after, a move of the sample X and Y coarse stages, so that when the move completes the sample piezo stages are in the center of their physical and servo control voltage range [14].

Most of the data collection processes are constructed from two basic scan types: Point scans, i.e., measurement of a single pixel, and fly scans, i.e., measurement of a line of pixels at a time. Currently most data is collected as ZP fly scanned images with OSA Y tracking. This acquisition process is described in greater detail below.

The user defines several input parameters, including the area to be imaged, the number of pixels to sample the area, and the dwell time at each pixel. This information is used to calculate a constant velocity at which the ZP X piezo stage will need to move at to satisfy the input parameters. The ZP X stage is moved outside the image area to a start location, and a marker position is defined in the E-712 at the border of the area to be imaged. The distance between the start position and the marker should be enough so that the ZP X stage can accelerate and reach the constant velocity by the time the stage arrives at the marker position. A counter input task is created for the PCIe-6351 to count its integrated scalar function, fed by the PMT detector, for a time equal to the dwell time per pixel multiplied by the number of pixels in the scan line. Finally, a task to generate a pulse train to gate the counter input task is also sent to the PCIe-6351. The time elapsed between two pulses is equal to the specified dwell time, and the number of pulses is equal to the number of pixels in the scan line.

When the scan is initiated and the ZP X stage passes over the marker, the E-712 produces a trigger pulse seen by the PCIe-6351 which initiates the counter and pulse train tasks. The counter immediately begins producing a table of the scalar value versus time, which is transformed into counts per pixel using the information from the pulse train. The scan line data is sent to a network drive and appears in a graphical plot in pySTXM. As the counter and pulse train tasks expire, the stage continues to move to a target position beyond the scan line so that constant velocity is maintained over the desired image area. When the stage reaches the target position, the previous marker position is cleared, and the stage returns to a start position that corresponds to the next line of the image. The ZP Y and OSA Y stages take a step, and the whole process is repeated for the number of lines required to cover the user defined image area.

A significant increase in piezo scanning speed relative to the ambient-STXM was achieved by employing the four onboard programmable waveform generators (FPGAs) in the E-712. The waveform generators execute the point or fly scan motion profiles with 5 µs time resolution, and are called repeatedly to form a raster scan. During execution of the move profiles, the E-712 also measures the following error over the entire motion profile using the interferometer. A proprietary control algorithm in the E-712, Dynamic Digital Linearization (DDL), computes a correction table that is applied to the next line. Over a few lines the tracking error is reduced to negligible levels. Once a scan is initiated, the move profiles and following error suppression are all performed in hardware, greatly reducing communication overhead. The correction tables are saved in a growing database, so that the DDL measurement doesn't need to be repeated for the same scan configuration, and the system learns as users fill in more scan possibilities.

**2.6 Vacuum**

The vacuum system was designed to achieve $10^{-8}$ Torr, with two priorities: 1) low partial pressure of water, which leads to ice formation on cryo-cooled samples [24,42], and 2) low partial pressure of hydrocarbons, which lead to carbon contamination on samples, and also on the vacuum window, ZP, OSA and detectors [58,59]. Standard ultra-high vacuum (UHV, $<1.0 \times 10^{-8}$ Torr) concepts are incorporated throughout the design such as venting trapped volumes and avoiding bolting large flat surfaces of internal components together. All stages and purchased internal components were UHV rated. Carbon containing materials were reduced or avoided wherever possible. For example, the vacuum chamber was fabricated in low carbon 304L stainless steel rather than 304, the electrical connectors sourced were ceramic rather than polyether ether ketone (PEEK), and a glass light pipe was sourced for the PMT detector rather than a traditional plastic piece [45]. The cryo-STXM is the first STXM designed to be compatible with in-situ plasma cleaning (GV10x, ibss Group). These units generate a reactive gas that flows through the vacuum chamber, decomposing hydrocarbons into smaller, more volatile molecules, which are then pumped out of the system [60]. UHV perfluoropolyether stage lubricants were chosen over hydrocarbon-based lubricants for greater resistance to plasma degradation. For wire insulation, Teflon® was preferred over Kapton® for its greater resistance to plasma [61] and lower water absorption [62]. Silver, which forms black particles in an oxidizing plasma, was avoided in vacuum or substituted with gold.

A schematic of the vacuum system is presented in **Supplemental Figure 2**. A re-entrant 100 nm thick silicon nitride window up to 1.5 mm × 1.5 mm separates the vacuum systems of the two STXMs. The system is pumped by one 700 L s$^{-1}$ turbopump directly mounted to the chamber baseplate with a gate valve and backed by a 4.3 L s$^{-1}$ scroll pump. A 100 mL capacity LN$_2$ cold finger/cryo-pump/anti-contaminator [42] is also installed. The majority of the flanges on the vacuum chamber are ConFlat®. Pumped double Viton® o-ring seals were used in cases where it was not practical to use ConFlats (e.g., the flange was too large or will be opened regularly). The chamber has 5.9 linear meters of o-ring seal, of which 5.2 meters are double o-ring. The goniometer has an integrated load lock system for introduction of samples without breaking vacuum. The load lock and all double o-rings are pumped by a common 1.8 L s$^{-1}$ scroll pump. In conventional mode, samples are currently exchanged through a port in the top of the chamber using forceps after breaking vacuum. All stainless steel, aluminum and bronze in vacuum was electropolished, including all hardware, brackets, and the entire chamber. After UHV cleaning [63], the empty chamber, gaskets, and all internal parts were baked in groups near their respective maximum temperatures at <10$^{-6}$ Torr before final assembly. The design incorporates infrared lamps to bake out the assembled instrument to a maximum internal temperature of 353 K, the limit of the sample and ZP piezo stages.

## 3. Results

### 3.1 Vibration and position stability

The difference in position of the ZP and sample in X and Y was measured under various conditions using the interferometer, and the results are presented in **Table I**. All measurements were made at a sampling rate of 20 kHz while holding position for 1 s with the turbopump and roughing pumps turned on.

**Table I**: Vibration measurements of the ZP relative to the sample under passive and active damping conditions.

| Conditions | | Displacement | |
|---|---|---|---|
| Damping pads | Feedback loop | X (nm RMS) | Y (nm RMS) |
| No | Off | 48.2 | 80.3 |
| Yes | Off | 13.5 | 17.3 |
| Yes | On | 0.7 | 0.6 |

Addition of the damping pads decreased vibration amplitudes by roughly 75% compared to a direct coupling of the block to the floor via four 10 cm × 10 cm × 10 cm thick wall steel square tubes. With the damping pads in place, engaging the interferometer feedback loop resulted in a further reduction, typically 95%, to sub-1 nm RMS levels. A Fourier analysis revealed that the majority of residual vibration with the damping pads and interferometer feedback in place lies in the regions 20-35 Hz and 80-100 Hz for both X and Y, with no features greater than 0.01 nm peak amplitude above 110 Hz. A feature in X and Y at 50 Hz with 0.1 nm peak amplitude disappeared when the pumps were turned off.

**3.2 Thermal stability**

The thermal stability of the Gatan 630 cryo sample holder was measured at the sample position while loaded in the cryo-STXM at $10^{-8}$ Torr, and the data is presented in **Figure 5**.

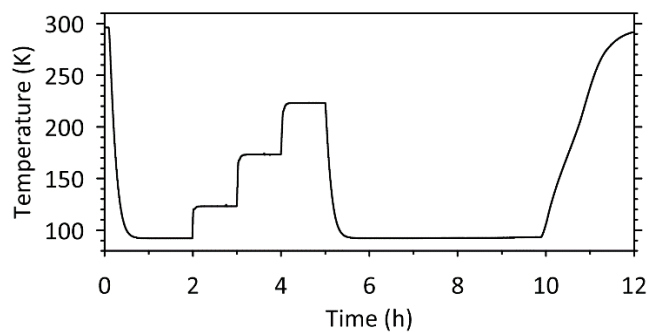

**Figure 5**: Temperature profile of the Gatan 630 sample holder while loaded in the cryo-STXM at $10^{-8}$ Torr.

Starting from room temperature, the integrated $LN_2$ dewar was filled at t = 0:05 h, and immediately the sample temperature began to drop, reaching a base temperature of 92.2 K at 0:55 h. This temperature can be held in the STXM for days so long as the dewar is periodically filled. The sample temperature was then increased in steps using the SmartSet 900 controller, which activates a heating circuit along a thermally conductive rod inside the sample holder in an evacuated space between the dewar and the sample. The temperature was held at 123 K, 173 K and 223 K with an accuracy of 0.1 K. Holding above 92 K is accompanied with increased $LN_2$ consumption. After holding at 223 K, the heating circuit was turned off and the temperature began falling. Upon reaching 92 K, the dewar was filled a final time at 5:35 h and the sample remained below 93 K until 9:40 h.

**3.3 Vacuum**

A typical pumpdown curve for the non-baked out system in goniometer mode is presented in **Figure 6**. During this cycle, the interferometer signal strength was observed to be stable; no adjustments were required.

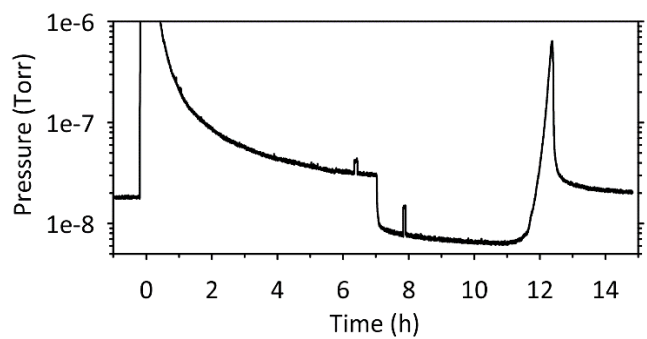

**Figure 6**: Pumpdown curve in goniometer mode, no bake out.

Initially the system was at $1.8 \times 10^{-8}$ Torr, the typical base pressure for goniometer mode. This level was achieved after running the turbopump and double o-ring pump for one week, with several samples introduced by load lock. The chamber was vented to 760 Torr using dry nitrogen, and pumping began at t = 0:00 h. The pressure dropped below $1.0 \times 10^{-7}$ Torr at 1:50 h. At 6:20 h, the double o-ring gaps were vented to atmosphere for several minutes, which resulted in a pressure increase from $3.0 \times 10^{-8}$ to $4.0 \times 10^{-8}$ Torr, or 33%. When the o-ring pumping was valved on, the pressure returned to $3.0 \times 10^{-8}$ Torr and continued dropping. The $LN_2$ cold finger was filled at 7:00 h, causing the pressure to drop more than half an order of magnitude into the $10^{-9}$ Torr or UHV region. Cooling the sample in-situ by filling the $LN_2$ dewar of the Gatan 630 did not lead to a significant drop in pressure as the surface area in vacuum that is cooled is only about 1 cm$^2$.

At 7:50 h, the double o-ring gaps were again vented to atmosphere for several minutes, which resulted in a pressure increase from $7.7 \times 10^{-9}$ to $1.5 \times 10^{-8}$ Torr, or 95%. The o-ring pumping was valved on again, and the system pressure returned to $7.7 \times 10^{-9}$ Torr and continued dropping. These results show that pumping the double o-ring gaps is increasingly effective at reduced pressures, where small leaks and permeation through the o-rings is a greater contribution to the total gas load. The double o-ring pumping and cold finger together allow this system with extensive o-ring seals to reach UHV without a bake out. The pressure reached a minimum value of $6.4 \times 10^{-9}$ Torr about 4 h after filling the cold finger with no refill. The pressure began to increase as the cold finger warmed to room temperature and captured species desorbed. At 14:30 h the system had reached $2.0 \times 10^{-8}$ Torr again.

The bake out lamps and plasma cleaner have also been commissioned. These systems could be activated to improve the vacuum in the future. With a bake out holding at 333 K measured internally for 8 days, the base pressure without the cold finger was $7.7 \times 10^{-9}$ Torr. Experiments with the plasma cleaner were performed using pure hydrogen gas, 90 W radio frequency power, and a chamber pressure of $2.3 \times 10^{-2}$ Torr. The in-vacuum components underwent several 10 minute exposures. This had no noticeable effect on the system (wiring, stages, PMT detector, etc.), while candle soot on a glass slide representing carbon contamination inside the chamber was partly removed. While these devices are now common on electron microscopes, to our knowledge this is the first application on a STXM. In addition to cleaning, the reactive gas presents an interesting sample environment not previously available for STXM. Further experiments quantifying the effects of the plasma cleaner are planned.

### 3.4 Spatial resolution
#### 3.4.1 Conventional mode

A gold on silicon nitride test pattern was imaged at room temperature and a pressure of $10^{-2}$ Torr to demonstrate the spatial resolution of the microscope in conventional mode. A transmission image of the test pattern is presented in **Figure 7**.

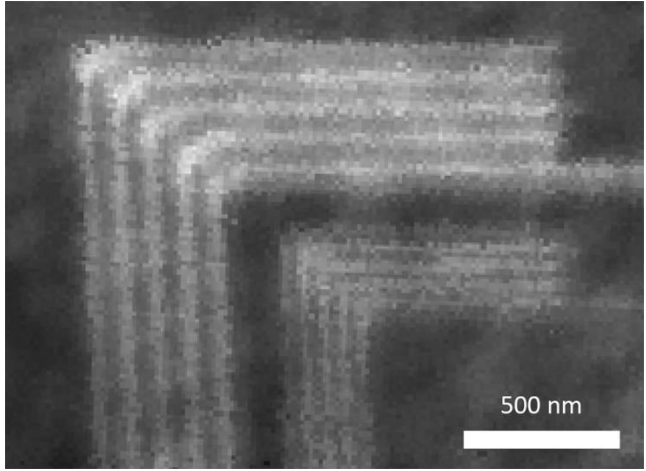

**Figure 7**: X-ray transmission image at 395.0 eV of a resolution test pattern with 50 nm and 30 nm half pitch elbows, acquired at 296 K and $10^{-2}$ Torr.

This image was acquired at 395.0 eV, using sample point by point scanning, and exit slits set to 30 µm × 30 µm. The 50 nm and 30 nm half pitch features are resolved. With regards to imaging speed, we define the efficiency of data acquisition as the sum of the number of pixels per image, multiplied by the dwell time per pixel each image, which is then divided by the actual time elapsed executing the scan. For example, the Fig. 7 uncropped image of 200 × 200 pixels (15 nm pixel size) with 1 ms dwell time per pixel took 126 s to acquire, giving an efficiency of (200 × 200 × 1 ms)/(126 s) = 31.7%.

### 3.4.2 Goniometer mode

The 35 nm half pitch outermost zones of a gold on silicon nitride ZP were imaged at a temperature of 92 K and a pressure of $10^{-9}$ Torr to demonstrate the spatial resolution of the microscope in goniometer mode under cryo measurement conditions. A transmission image of the ZP is presented in **Figure 8**.

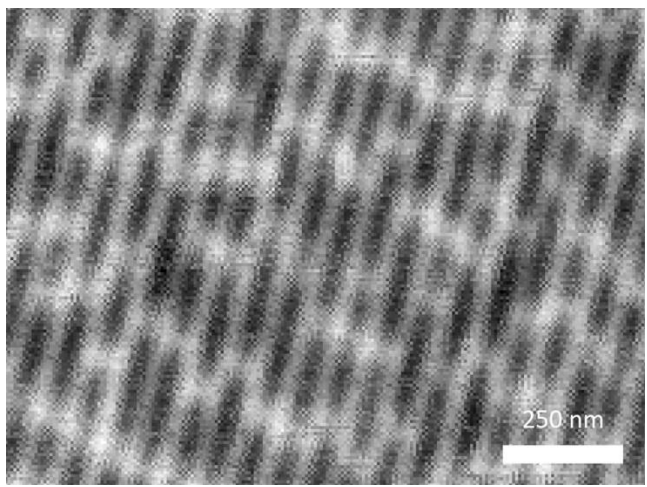

**Figure 8**: X-ray transmission image at 694.0 eV of 35 nm outermost zones of a gold ZP, acquired at 92 K and $10^{-9}$ Torr.

This image was acquired at 694.0 eV, using ZP fly scanning, and exit slits set to 50 µm × 50 µm. The 35 nm half pitch zones are resolved. A Fourier ring correlation analysis [64,65] was performed on this image, revealing a half pitch spatial resolution of 31 nm using the 0.5 threshold, and 29 nm using the half-bit threshold (**Supplemental Figure 3**). The uncropped image of 286 × 286 pixels (7 nm pixel size) with 5 ms dwell time per pixel took 426 s to acquire, an efficiency of 96.0%.

### 3.5 Spectromicroscopy

An image stack [6] covering the F 1s absorption region was collected from a thin section of a polymer electrolyte membrane (PEM) hydrogen fuel cell. The sample underwent a microtoming procedure described elsewhere [66]. The stack was collected at 92 K and $10^{-8}$ Torr in goniometer mode. It consisted of 92 images covering 680 – 740 eV, with 180 × 230 pixels (200 nm pixel size) and 1 ms dwell time per pixel. This stack took 5071 s to acquire, an efficiency of 75.1%. Results derived from the stack are presented in **Figure 9**.

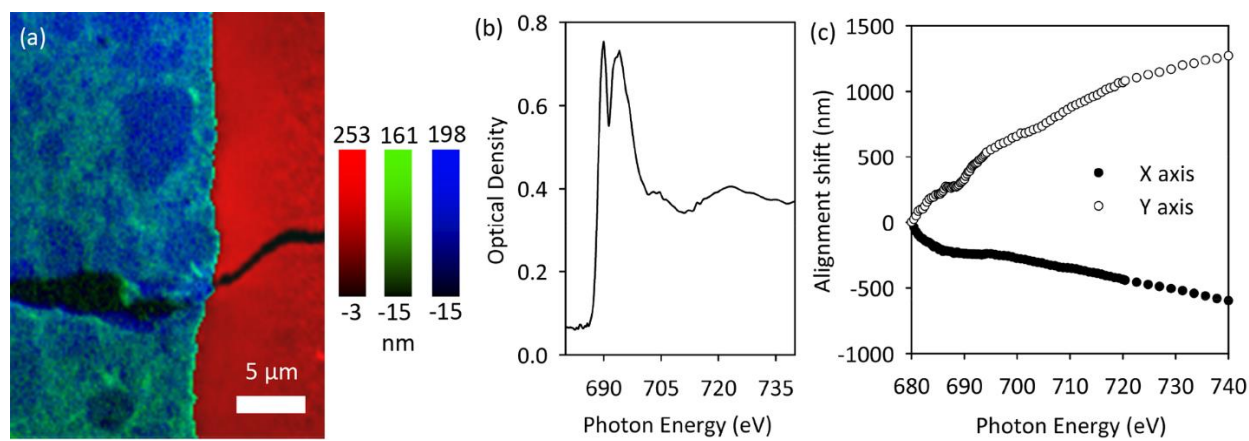

**Figure 9**. Results derived from a F 1s stack of a PEM fuel cell sample. (a) Color coded composite of thickness maps in nanometers of compounds in the sample: PFSA in the membrane in red, PFSA in the cathode in green, and carbon support in the cathode in blue. (b) F 1s XANES spectrum derived from the membrane region. (c) The post-exposure alignment file for the stack of 92 images.

Fig. 9 a) is a color composite image of quantitative component thickness maps in nanometers derived from the stack data. The regions of the PEM fuel cell were initially assigned by comparing X-ray transmission images in the stack to optical microscope images taken during sample preparation (**Supplemental Figure 4**). The relatively uniform region on the right of Fig. 9 a) is the membrane itself, which is polyfluorosulfonic acid (PFSA). An X-ray transmission spectrum was extracted from this region, and then converted to optical density (OD, Fig. 9 b) using an $I_0$ transmission spectrum extracted from the horizontal crack in the sample, an artefact of the sample preparation. This F 1s XANES spectrum with major peaks at 689.9 and 694.0 eV is that of PFSA [67], and is correct for this region of this sample [66]. A thickness map of the membrane (in red) was derived by masking this region from the rest of the stack and dividing the OD spectrum from each pixel in the masked region by an OD nm$^{-1}$ spectrum of PFSA. The cathode region of the fuel cell is captured in the left of Fig. 9 a). This region is known to be a complex mixture of PFSA and a graphitic carbon support decorated with catalytic platinum nanoparticles [30,66]. A thickness map of PFSA in the cathode (in green) was created from two OD images in the stack: the image at 694.0 eV (peak of PFSA), minus the OD image at 680.0 eV (pre-F 1s). The resulting image was then divided by a constant of 0.0040 OD nm$^{-1}$ [66], and the membrane region was removed by masking. Lastly, a map of the non-fluorine material (carbon support and catalyst nanoparticles, in blue) in the cathode was created by subtracting the contribution of PFSA in the cathode, the only fluorine containing component in the sample, from the OD image at 680.0 eV. The contribution of PFSA was determined by multiplying the thickness map of PFSA in the cathode by a scaling factor, derived from the OD nm$^{-1}$ spectrum of PFSA [66]. This gave carbon support thickness values that were in agreement with the measured thickness of the membrane.

Overall, the spectromicroscopy data compiled in Fig. 9 a) and b) reveal that the membrane region consists of pure PFSA and is of uniform thickness, while the cathode region consists of a relatively uniform distribution of platinum nanoparticle coated carbon support with a non-uniform PFSA distribution, with domains that are significantly depleted in PFSA. This data, collected at 92 K and 10$^{-8}$ Torr, is completely consistent with data obtained from similar PEM samples on other STXMs, including the ambient-STXM [30,66,67]. It is also interesting with regard to the limits of ZP scanning, as the uncropped stack was of a relatively large area (36 µm × 46 µm) and displayed no loss of spatial resolution out to the edges of the images, albeit at the modest pixel spacing of 200 nm.

Although STXMs have incorporated interferometers for nearly 20 years to suppress image drift during acquisition (§2.5), in practice some drift is usually observed. STXM data analysis software such as aXis2000 and MANTiS incorporate post-image cross correlation algorithms to correct for this [6]. A plot of the alignment file that was applied to the stack, which is a record of drift, is presented in Fig. 9 c). This drift is partly due to misalignment of the ZP interferometer mirrors (the ZP translated 290 µm in Z to maintain focus over the 60 eV stack) and the STXM chamber which are adjustable and can be improved, and partly due to random thermal motion over the nearly hour and a half acquisition time. The trends from several stacks taken the same day were consistent, suggesting mirror and/or chamber misalignment played the larger role. In any case, the amount of drift and the trends would be typical for the ambient-STXM under similar conditions and are easily corrected by established methods. The performance is remarkable considering that when operating in goniometer mode, the set of interferometer mirrors which represent the sample motion are located far from the most ideal position directly on the tip of the sample holder.

### 3.6 Cryo-suppression of radiation damage

Although the data in §3.2 indicate that a sample temperature as low as 92 K can be achieved, it was still crucial to verify that cryo-cooling had the intended effect on radiation damage. After all, cryo-suppression of radiation damage to enable artefact-free high dose STXM measurements was the main motivation for developing the instrument. The verification experiment can be rather simple: Given constant exposure conditions/dose, a sample exposed at room temperature should show more damage compared to a sample exposed at near LN2 temperature. In addition, some portion of cryo-suppressed damage should be revealed when the exposed cryo-cooled sample is allowed to warm up to room temperature [31-34].

A 100 ±14 nm thick region of a PEM fuel cell thin section sample was used for the verification experiment, performed at $10^{-8}$ Torr. A 5 µm × 5 µm area of the sample containing both membrane and cathode regions was exposed at 296 K to 694.0 eV X-rays for 220 ms per 50 nm pixel. A dose of 72 MGy was calculated for the uniform membrane region using a dose calculation method described elsewhere [68]. This area was imaged at 694.0 eV immediately after the exposure (**Figure 10 a**). The exposed area was visible in the image as a lighter square, due to an increase in transmission in the membrane region. 694.0 eV corresponds to the specific energy of a strong F 1s → $\sigma^*_{C-F}$ absorption peak (Fig. 9 b) [67], combined with some non-specific photoabsorption for all other elements in the material. An increase in transmission is therefore strongly indicative of a loss of C-F bonds and/or mass in the exposure area due to the dose.

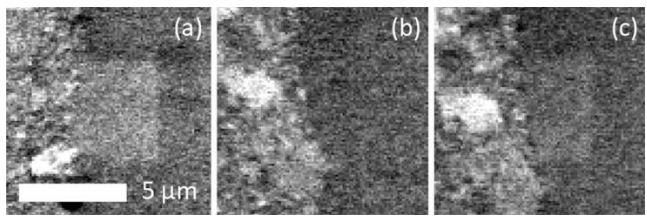

**Figure 10**. X-ray transmission images at 694.0 eV of 100 ±14 nm thick PEM fuel cell sample. The membrane region is on the right and the cathode is on the left for all images, which are all on the same spatial scale. (a) Image of a sample region immediately after X-ray exposure at 296 K. (b) Image of a different sample region immediately after X-ray exposure at 95 K, using otherwise identical exposure conditions as a). (c) Same region as b), imaged after allowing the sample to warm up to 296 K.

The sample was then cooled in vacuum to 95 K by filling the Gatan 630 dewar with $LN_2$. The exact same exposure/dose was performed at 95 K on a new 5 µm × 5 µm area of the sample. This area was imaged at 694.0 eV immediately after exposure while maintaining 95 K (Fig. 10 b). No increase in transmission was observed for the exposed region. Finally, with the sample remaining in vacuum, LN$_2$ was evacuated from the dewar and the heating circuit was activated until the sample temperature reached 296 K. The region of the sample that was exposed at 95 K was again imaged at 694.0 eV (Fig. 10 c), 45 minutes after the cryo-exposure completed. The exposed region was now visible in the 296 K image as a lighter square, with an increase in transmission which was the same within error as that measured for the exposure performed at 296 K. These observations of cryo-suppression and later revealing of radiation damage are consistent with previous STXM investigations on different samples [33,34]. Spectroscopic measurements were also collected and will be presented elsewhere.

### 3.7 Cryo-spectrotomography

A PEM fuel cell thin section sample was used to demonstrate cryo-tomographic performance. All data was collected at a sample temperature of 92 K and pressure of $10^{-8}$ Torr. X-ray transmission images at 684.0 and 694.0 eV were collected at θ angles from -60° to +60° in 5° steps for a total of 50 images. Each uncropped image was 240 × 1000 pixels (50 nm pixel size) and was recorded using 1 ms dwell time per pixel. This data set took 14951 s to acquire, an efficiency of 80.3%. The acquisition executed automatically in pySTXM, with no user input after setting up the scan. A simple linear refocus was applied to ZP Z over the rotation range, and the goniometer motors were stationary except for changing θ for each tilt angle. By tracking a region of interest in the images, the deviation from eucentric positioning over the -60° to +60° θ range was found to be 2 µm in X, 42 µm in Y, and 150 µm in Z.

Quantitative spectrotomographs of PFSA and non-fluorine material (carbon support and catalyst nanoparticles) were created in the following manner: For each angle, the two transmission images were aligned, then converted to OD. The OD image at 684.0 eV (pre-F 1s) was subtracted from the OD image at 694.0 eV (peak of PFSA), creating PFSA OD maps. OD maps of the carbon support at each angle were created by subtracting the PFSA map, multiplied by a scaling factor, from the OD image at 684.0 eV to correct for PFSA absorption at 684.0 eV (§3.5). The PFSA and carbon support maps at each angle (**Supplemental Figure 5**) were manually cropped and aligned individually, and then finely aligned as a group using cross correlation algorithms in aXis2000. 3D reconstructions were performed using a compressed sensing algorithm [69] in MANTiS, with a beta parameter of 0.1, 50 iterations, and the initial thickness parameter set to the width derived from the average OD. Avizo [70] renderings of the reconstructed data are presented in **Figure 11** and as a **movie** in the supplemental information.

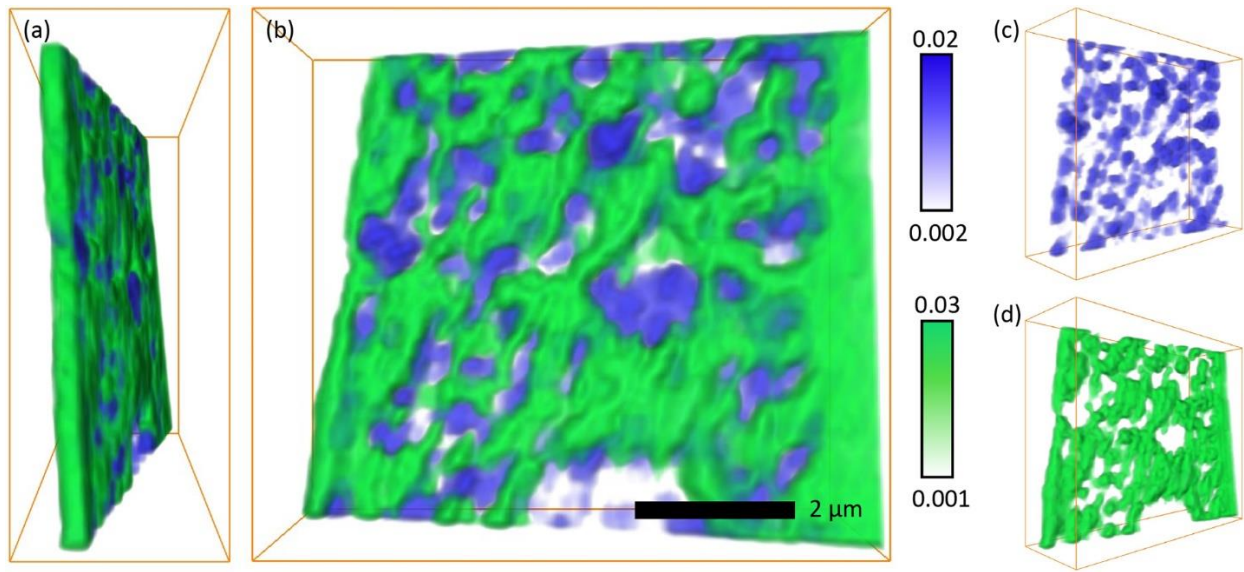

**Figure 11**: 4D volume rendered images of a PEM fuel cell thin section, with PFSA in green and carbon support in blue, (a) Z-Y view, and (b) X-Y view. Individual spectrotomographs of (c) carbon support and (d) PFSA. Otsu threshold [71] values were 0.001 in the range of 0-0.03 for PFSA, 0.002 in the range 0-0.02 for carbon support.

The distribution of PFSA and carbon support, and the pore space and connectivity, are critical to the optimization of efficient and low cost PEM fuel cells [30], and this type of data is valuable feedback for both the formulation and the manufacturing of fuel cell stacks. The cryo-tomographic data set contains a more accurate description of the 3D distribution compared with the same material measured in 2D (Fig. 9a). Slices along the thickness direction Z are presented as **Supplemental Figure 6**. The shape of the particles changes from slice to slice, capturing the true 3D structure of the PFSA and carbon support. In general, the carbon support particles are surrounded by PFSA with good colocalization. But there are significant volumes where PFSA is relatively far from carbon particles, indicating that the amount of PFSA and the porosity of this sample may not be optimum.

The half pitch spatial resolution of this tomographic imaging demonstration is estimated to be 50 nm in the X and Y planes and 75 nm in the Z plane by evaluating several line intensity profiles over high contrast features [5]. The resolution was limited by the choice of 50 nm pixel size, which was in turn related to the desire to present a fully automated cryo-spectrotomography measurement. A better result could have been achieved if a finer pixel spacing was chosen (Fig. 8). This would be practical if the field of view was reduced by a) manually reorienting and refocusing at each tilt angle, or ideally b) improving eucentricity by automated adjustment of the X Y and Z sample positions during the scan, and optimization of the sample mounting procedure.

**4. Discussion**

In this report we have demonstrated a level of performance and functionality that will be a baseline for continuous development. Several visiting user groups have now collected data using the instrument, and there are already avenues for improvement in progress. Implementation will be primarily driven by demand from the SM user community.

**4.1 Harder sample conditions, better data and new data**

The motivation for the cryo-STXM was to reduce radiation damage and sample contamination for high dose measurements. Cooling to even lower temperature may provide greater damage reduction in certain situations [72,73], however it would be an even greater technical challenge with diminishing returns. The same would be true for pursuing even higher vacuum with regard to contamination. These activities are not currently planned. Provisions for an inboard-side-entry sample load lock for conventional mode are available. Although this would provide higher average vacuum in conventional mode, the main goal would be high throughput remote access. This development is currently on hold as the demand has been strongest for goniometer mode, particularly cryo experiments.

Although all cryo experiments presented were performed by loading the sample at room temperature and then cooling it in vacuum, the current system is fully capable of loading a previously plunge frozen sample [44]. Cryo sample preparation infrastructure is currently being added to the SM facility as a high priority. The Gatan 630 sample holder used throughout this report enables cryo-tomography as well as temperature dependent experiments. However, the goniometer readily accepts any TEM sample holder in the JEOL 2010 pattern. There is a wide range of commercially available sample holders which enable humidity, electrochemical, high temperature, gas flow, and other in-situ experiments [74]. By choosing to use TEM sample holders for this STXM, we leverage all of the developments in TEM in-situ set ups, and reduce barriers for correlative TEM-STXM experiments.

The pursuit of higher spatial resolution will be a continuing process. To our knowledge this is the first instrument in user operation that relies on the combination of zone plate scanning and a TEM goniometer as the sample holder, with seemingly non-optimal sample mirror locations for closing the interferometer feedback loop. For this debut, we only sought to achieve the level of spatial resolution that is now considered routine at most STXM facilities, but now with the added environmental conditions of 92 K sample temperature and $10^{-9}$ Torr. The results for the accuracy and vibration of the scanning mechanisms (§3.1) indicate that the stages are not limiting the spatial resolution, and better than 30 nm should be obtainable by using better ZPs and finer test patterns. These investigations are now underway. Enhanced spatial resolution can also be obtained by ptychography [26], and an area detector to enable ptychographic experiments is being procured. Other detection modes including sample current [23,24] and X-ray fluorescence [25] are also planned.

**4.2 Greater access**

Given the limited and highly regulated access to synchrotron beam time, a STXM coupled with a lab based soft X-ray source would greatly expand access and applications of the technique. But the requirements for the source are quite high. The scanning optical arrangement involving a ZP requires high spatial coherence to reach competitive spatial resolution [1], and the spatial coherence of current lab sources is low. This can be improved with spatial filtering at the cost of already limited flux. In addition, most applications of STXM involve XANES spectromicroscopy, so it would be critical for the source to be tunable with an energy resolving power (E/ΔE) of at least 3000 [1]. Put simply, the flux, coherence, and tunability of the best present lab sources is not competitive with synchrotron sources, and all operational soft X-ray STXMs we are aware of use a synchrotron beamline as the X-ray source [3]. Nevertheless, there have been attempts at a lab source based STXM [75], and we have made both the ambient- and cryo-STXMs operate on a lab source: a 405 nm optical laser introduced downstream of the SM beamline exit slits. By simply replacing the ZP with a conventional convex lens for visible light, the microscopes function as scanning transmission optical microscopes, and produce images independent of the beamline. Half pitch spatial resolution below 500 nm has been achieved (**Supplemental Figure 4**). This lab source mode has effectively increased access by shifting certain activities such as software development, stage tuning and alignment, and user training into long shutdown periods.

Improving the efficiency of data acquisition will be a continuing process towards increasing access. STXM data collection often involves automated sequences of multiple scans, such as image stacks, and most of these are now optimized in pySTXM. Optimization of automated tomography is currently underway. The reasons to automate and many of the concepts such as eucentric adjustment and image alignment are identical to TEM tomography [46]. The tolerances for autofocusing are challenging due to the limited depth of focus of high resolution ZP lenses [1,5]. A sinusoidal focus correction, and a θ-dependent focus adjustment to accommodate the tilt of the sample are being implemented.

The speed of some experiments, including ptychography and fluorescence in the ambient-STXM, are mainly limited by the available flux. These modes would immediately benefit from a brighter source such as a multi-bend achromat synchrotron. But for most experiments, the efficiency is limited by the scan speed. We have achieved a significant improvement in the efficiency of data acquisition relative to the ambient-STXM: on average, a scan on the cryo-STXM is complete in 60% of the time taken using the ambient-STXM, using identical input parameters (image area, pixel sampling, dwell time, number of energies, tilt angles, etc.). This is largely due to the ZP and sample piezo scans executing in the hardware (§2.5). Investigations into bi-directional scanning could incrementally improve the performance. More substantial increases in scan speed could involve emerging stage technology based on resonant scanning [76] rather than piezo flexure arrangements, or scanning the position and angle of the electron beam/source, rather than the samples or optics [75,77]. There are already conditions where the stages can scan too fast. The minimum pixel dwell time is usually limited to 1 ms due to limitations of the counting system or the finite decay time of the phosphor of the PMT detector, which can lead to a loss in spatial resolution at higher scan speeds [78]. Phosphors with shorter decay times are available, but typically come with a trade-off of decreased dose efficiency. Avalanche photodiodes provide gigahertz count rates but have low dose efficiency below 600 eV. In summary, the move toward sub-1 ms dwell times in STXM will be contingent on developments in source, shutter, stage motion and detector technology, and is likely to become more sensitive to specific types of samples and/or experiments.

## 5. Conclusions

A cryo scanning transmission X-ray microscope was developed for independent user operation. The microscope has demonstrated 30 nm half pitch spatial resolution in the soft X-ray region using a 25 nm outermost zone width zone plate, while maintaining a sample temperature of 92 K and a pressure of $10^{-9}$ Torr. XANES spectromicroscopy was demonstrated on hydrogen fuel cell microtomed sections under cryo conditions. The acquisition times were on average 60% of the CLS ambient-STXM while maintaining equivalent spatial and spectral resolution. Radiation damage was shown to be temporally suppressed under cryo conditions. The first demonstration STXM-based of cryo-spectrotomograpy in the soft X-ray region (<1 keV) was presented.

## 6. Supplementary Material

See supplementary material for a table of selected parameters of commercial stages utilized in the CLS cryo-STXM, images of various coordinated ZP-OSA scans, a simplified diagram of the vacuum system, a Fourier ring correlation analysis of the X-ray transmission image presented in Figure 8, a description and demonstration of the optical laser imaging mode, all PFSA and carbon support maps used for the tomographic reconstruction presented in Figure 11, slices along the Z thickness direction of the 4D data set presented in Figure 11, and a video of the Avizo rendering of the 4D data set presented in Figure 11.

## 7. Acknowledgements

and Jeurgen Stumper (Automotive Fuel Cell Cooperation) are thanked for providing funding as well as the automotive hydrogen fuel cell samples which were further prepared by Marcia West (McMaster University). Benjamin Watts (Swiss Light Source) is thanked for the use of NXstxm file format. Tolek Tyliszczak is thanked for his role as an external reviewer of the design. Robert Peters and Mirwais Aktary (Applied Nanotools) are thanked for their collaborative effort in producing the resolution test patterns and custom zone plates. Xiaohui Zhu (Shanghai University) is thanked for providing a Fourier ring correlation analysis program and assisting with the interpretation of those results. Yingshen Lu, James Dynes and Jigang Zhou (CLS) are thanked for maintaining the SM beamline.

The CLS cryo-STXM is a component of a CFI Leading Edge fund project, "Enhancing the Spectromicroscopy Beamline and Endstations at the Canadian Light Source", with financial support from Automotive Fuel Cell Cooperation, the Government of Saskatchewan, the Government of Ontario, and the Government of Alberta. Research described in this paper was performed at the Canadian Light Source, which is supported by the Canada Foundation for Innovation, Natural Sciences and Engineering Research Council of Canada, the University of Saskatchewan, the Government of Saskatchewan, Western Economic Diversification Canada, the National Research Council Canada, and the Canadian Institutes of Health Research.

## 8. References

[1] M. Howells, C. Jacobsen, and T. Warwick, "Principles and Applications of Zone Plate X-ray Microscopes" in P. W. Hawkes, J. C. Spence, eds. *Science of Microscopy* vol. 2, ch. 13 (Springer Science+Business Media: New York) 2007.

[2] H. Ade and A. P. Hitchcock, Polymer **49**, 643-675 (2008).

[3] A. P. Hitchcock, J. Elec. Spec. Relat. Phenom. **200**, 49-63 (2015).

[4] K. V. Kaznatcheev, C. Karunakaran, U. D. Lanke, S. G. Urquhart, M. Obst, and A. P. Hitchcock, Nucl. Inst. Meth. Phys. Res., Sect. A **582**, 96-99 (2007).

[5] D. Attwood, *Soft X-rays and Extreme Ultraviolet Radiation* (Cambridge University Press: Cambridge) 1999.

[6] C. Jacobsen, S. Wirick, G. Flynn, and C. Zimba, J. Microsc. **197**, 173-184 (2000).

[7] B. Gilbert, S. C. Fakra, T. Xia, S. Pokhrel, L. Mädler, and A. E. Nel, ACS Nano **6**, 4921-4930 (2012).

[8] J. Lehmann, D. Solomon, J. Kinyangi, L. Dathe, S. Wirick, and C. Jacobsen, Nat. Geosci. **1**, 238-242 (2008).

[9] R. C. Moffet, T. C. Rödel, S. T. Kelly, X. Y. Yu, G. T. Carroll, J. Fast, R. A. Zaveri, A. Laskin, and M. K. Gilles, Atmos. Chem. Phys. **13**, 10445-10459 (2013).

[10] A. F. G. Leontowich and A. P. Hitchcock, Microfluid. Nanofluid. **15**, 509-518 (2013).

[11] J. Lim, Y. Li, D. H. Alsem, H. So, S. C. Lee, P. Bai, D. A. Cogswell, X. Liu, N. Jin, Y. Yu, N. J. Salmon, D. A. Shapiro, M. Z. Bazant, T. Tyliszczak, and W. C. Chueh, Science **353**, 566-571 (2016).

[12] H. Rarback, J. M. Kenney, J. Kirz, M. R. Howells, P. Chang, P. J. Coane, R. Feder, P. J. Houzego, D. P. Kern, and D. Sayre, "Recent Results from the Stony Brook Scanning Microscope" in G. Schmahl, D. Rudolf, eds. *X-Ray Microscopy* pg. 203-216 (Springer: Berlin) 1984.

[13] H. Rarback, D. Shu, S. C. Feng, H. Ade, J. Kirz, I. McNulty, D. P. Kern, T. H. P. Chang, Y. Vladimirsky, N. Iskander, D. Attwood, K. McQuaid, and S. Rothman, Rev. Sci. Instrum. **59**, 52-59 (1988).

[14] A. L. D. Kilcoyne, T. Tyliszczak, W. F. Steele, S. Fakra, P. Hitchcock, K. Franck, E. Anderson, B. Harteneck, E. G. Rightor, G. E. Mitchell, A. P. Hitchcock, L. Yang, T. Warwick, and H. Ade, J. Synchrotron Rad. **10**, 125-136 (2003).

[15] D. Sayre, J. Kirz, R. Feder, D. M. Kim, and E. Spiller, Ultramicroscopy **2**, 337-349 (1977).

[16] E. G. Rightor, A. P. Hitchcock, H. Ade, R. D. Leapman, S. G. Urquhart, A. P. Smith, G. Mitchell, D. Fischer, H. J. Shin, and T. Warwick, J. Phys. Chem. B **101**, 1950-1960 (1997).

[17] A. P. Hitchcock, "Soft X-ray Imaging and Spectromicroscopy" in G.V. Tendeloo, D. V. Dyck, S. J. Pennycook, eds. *Handbook of Nanoscopy* vol. 2, ch. 22 (Wiley-VCH Verlag: Weinheim) 2012.

[18] W. S. Haddad, I. McNulty, J. E. Trebes, E. H. Anderson, R. A. Levesque, and L. Yang, Science **266**, 1213-1215 (1994).

[19] B. F. McEwen, K. H. Downing, and R. M. Glaeser, Ultramicroscopy **60**, 357-373 (1995).

[20] C. G. Schroer, M. Kuhlmann, T. F. Günzler, B. Lengeler, M. Richwin, B. Griesebock, D. Lützenkirchen-Hecht, R. Frahm, E. Ziegler, A. Mashayekhi, D. R. Haeffner, J. –D. Grunwaldt, and A. Baiker, Appl. Phys. Lett. **82**, 3360-3362 (2003).

[21] G. A. Johansson, T. Tyliszczak, G. E. Mitchell, M. H. Keefe, and A. P. Hitchcock, J. Synchrotron Rad. **14**, 395-402 (2007).

[22] G. Schmid, F. Zeitvogel, L. Hao, P. Ingino, W. Kuerner, J. J. Dynes, C. Karunakaran, J. Wang, Y. Lu, T. Ayers, C. Schietinger, A. P. Hitchcock, and M. Obst, Microsc. Microanal. **20**, 531-536 (2014).

[23] C. Hub, S. Wenzel, J. Raabe, H. Ade, and R. H. Fink, Rev. Sci. Instrum. **81**, 033704 (2010).

[24] C. Stahl, S. Ruoβ, M. Weigand, M. Bechtel, G. Schütz, and J. Albrecht, AIP Conf. Proc. **1696**, 020031 (2016).

[25] A. P. Hitchcock, M. Obst, J. Wang, Y. S. Lu, and T. Tyliszczak, Environ. Sci. Tech. **46**, 2821-2829 (2012).

[26] D. A. Shapiro, Y. S. Yu, T. Tyliszczak, J. Cabana, R. Celestre, W. Chao, K. Kaznatcheev, A. L. D. Kilcoyne, F. Maia, S. Marchesini, Y. S. Meng, T. Warwick, L. L. Yang, and H. A. Padmore, Nat. Photon. **8**, 765-769 (2014).

[27] J. Kirz and C. Jacobsen, J. Phys.: Conf. Ser. **186**, 012001 (2009).

[28] M. R. Howells, T. Beetz, H. N. Chapman, C. Cui, J. M. Holton, C. J. Jacobsen, J. Kirz, E. Lima, S. Marchesini, H. Miao, D. Sayre, D. A. Shapiro, J. C. H. Spence, and D. Starodub, J. Elec. Spec. Relat. Phenom. **170**, 4-12 (2009).

[29] T. Coffey, S. G. Urquhart, and H. Ade, J. Elec. Spec. Rel. Phenom. **122**, 65-78 (2002).

[30] J. Wu, L. G. A. Melo, X. Zhu, M. M. West, V. Berejnov, D. Susac, J. Stumper, and A. P. Hitchcock, J. Power Sources **381**, 72-83 (2018).

[31] E. F. Garman and R. L. Owen, Acta. Cryst. **D62**, 32-47 (2006).

[32] R. F. Egerton, Ultramicroscopy **5**, 521-523 (1980).

[33] T. Beetz and C. Jacobsen, J. Synchrotron Rad. **10**, 280-283 (2003).

[34] J. Maser, A. Osanna, Y. Wang, C. Jacobsen, J. Kirz, S. Spector, B. Winn, and D. Tennant, J. Microsc. **197**, 68-79 (2000).

[35] G. Schneider, P. Guttmann, S. Rehbein, S. Werner, and R. Follath, J. Struct. Biol. **177**, 212-223 (2012).

[36] H. M. Hertz, O. V. Hofsten, M. Bertilson, U. Vogt, A. Holmberg, J. Reinspach, D. Martz, M. Selin, A. E. Christakou, J. Jerlström-Hultqvist, and S. Svärd, J. Struct. Biol. **177**, 267-272 (2012).

[37] M. A. Le Gros, G. McDermott, B. P. Cinquin, E. A. Smith, M. Do, W. L. Chao, P. P. Naulleau, and C. A. Larabell, J. Synchrotron Rad. **21**, 1370-1377 (2014).

[38] J. Maser, H. Chapman, C. Jacobsen, A. Kalinovsky, J. Kirz, A. Osanna, S. Spector, S. Wang, B. Winn, S. Wirick, and X. Zhang, Proc. SPIE **2516**, 78-89 (1995).

[39] J. Maser, C. Jacobsen, J. Kirz, A. Osanna, S. Spector, S. Wang, and J. Warnking, "Development of a Cryo Scanning Transmission X-Ray Microscope at the NSLS" in J. Thieme, G. Schmahl, D. Rudolph, E. Umbach, eds. *X-Ray Microscopy and Spectromicroscopy* pg. I-35 - I-44 (Springer-Verlag: Berlin Heidelberg) 1998.

[40] B. D. A. Levin, M. J. Zachman, J. G. Werner, R. Sahore, K. X. Nguyen, Y. Han, B. Xie, L. Ma, L. A. Archer, E. P. Giannelis, U. Wiesner, L. F. Kourkoutis, and D. A. Muller, Microsc. Microanal. **23**, 155-162 (2017).

[41] B. D. Cullity, *Elements of X-ray Diffraction* 2$^{nd}$ ed. pg. 135-139 (Addison-Wesley: Don Mills) 1978.

[42] X. Huang, H. Miao, J. Nelson, J. Turner, J. Steinbrener, D. Shapiro, J. Kirz, and C. Jacobsen, Nucl. Instrum. Meth. Phys. Res., Sect. A **638**, 171-175 (2011).

[43] T. Beetz, M. R. Howells, C. Jacobsen, C. C. Kao, J. Kirz, E. Lima, T. O. Mentes, H. Miao, C. Sanchez-Hanke, D. Sayre, and D. Shapiro, Nucl. Instrum. Meth. Phys. Res., Sect. A **545**, 459-468 (2005).

[44] Model 630 High Tilt Tomography, Instruction Manual, Gatan Item. No. 630.40000 (Nov. 2001). Model 900 SmartSet Cold Stage Controller, Instruction Manual, Gatan Part No. 900.40000, Rev. 10 (Feb. 2013).

[45] A. F. G. Leontowich, D. M. Taylor, J. Wang, C. N. Regier, T. Z. Regier, R. Berg, D. Beauregard, J. J. Dynes, C. Senger, J. Swirsky, C. Karunakaran, A. P. Hitchcock, and S. G. Urquhart, J. Phys.: Conf. Ser. **849**, 012045 (2017).

[46] A. J. Koster, H. Chen, J. W. Sedat, and D. A. Agard, Ultramicroscopy **46**, 207-227 (1992).

[47] M. Weigand, BESSY II, private communication.

[48] C. N. Regier, A. F. G. Leontowich, and D. M. Taylor, Proc. MEDSI2016, THAA02, 381-386 (2016).

[49] W. Thur, R. DeMarco, B. Baldock, and K. Rex, Proc. 5[th] Intl. Workshop on Accel. Align., 045 (1997).

[50] B. Samali and K. C. S. Kwok, Eng. Struct. **17**, 639-654 (1995).

[51] D. Mangra, S. Sharma, and J. Jendrzejczyk, Rev. Sci. Instrum. **67**, 3374 (1996).

[52] J. W. Li, E. Matias, N. Chen, C. -Y. Kim, J. Wang, J. Gorin, F. He, P. Thorpe, Y. Lu, W. F. Chen, P. Grochulski, X. B. Chen, and W. J. Zhang, J. Synchrotron Rad. **18**, 109-116 (2011).

[53] Experimental Physics and Industrial Control System, www.aps.anl.gov/epics/, accessed May 22, 2018.

[54] M. Newville, PyEpics: Epics Channel Access for Phython, http://cars9.uchicago.edu/software/python/pyepics3/index.html, accessed May 22, 2018.

[55] B. Watts and J. Raabe, AIP Conf. Proc. **1696**, 020042 (2016).

[56] A. P. Hitchcock, aXis2000, http://unicorn.mcmaster.ca/aXis2000.html, accessed May 22, 2018.

[57] M. Lerotic, R. Mak, S. Wirick, F. Meirer, and C. Jacobsen, J. Synchrotron Rad. **21**, 1206-1212 (2014).

[58] A. F. G. Leontowich and A. P. Hitchcock, J. Vac. Sci. Technol. B **30**, 030601 (2012).

[59] B. Watts, N. Pilet, B. Sarafimov, K. Witte, and J. Raabe, J. Instrumentation **13**, C04001 (2018).

[60] E. Pellegrin, I. Šics, J. Reyes-Herrera, C. P. Sempere, J. J. L. Alcolea, M. Langlois, J. F. Rodriguez, and V. Carlino, J. Synchrotron Rad. **21**, 300-314 (2014).

[61] K. K. De Groh, B. A. Banks, C. E. McCarthy, R. N. Rucker, L. M. Roberts, and L. A. Berger, High Perf. Polymers **20**, 388-409 (2008).

[62] M. Sampson, NASA Parts Selection List, https://nepp.nasa.gov/npsl/Wire/insulation_guide.htm, accessed May 22, 2018.

[63] Vacuum Component Cleaning Technical Procedure, Canadian Light Source document 8.7.33.1. Rev. 3.

[64] N. Banterle, K. H. Bui, E. A. Lemke, and M. Beck, J. Struct. Biol. **183**, 363-367 (2013).

[65] R. P. J. Nieuwenhuizen, K. A. Lidke, M. Bates, D. L. Puig, D. Grünwald, S. Stallinga, and B. Rieger, Nat. Meth. **10**, 557-562 (2013).

[66] A. P. Hitchcock, V. Berejnov, V. Lee, M. West, V. Colbow, M. Dutta, and S. Wessel, J. Power Sources **266**, 66-78 (2014).

[67] Z. B. Yan, R. Hayes, L. G. A. Melo, G. R. Goward, and A. P. Hitchcock, J. Phys. Chem. C **122**, 3233-3244 (2018).

[68] A. F. G. Leontowich, A. P. Hitchcock, T. Tyliszczak, M. Weigand, J. Wang, and C. Karunakaran, J. Synchrotron Rad. **19**, 976-987 (2012).

[69] E. Y. Sidky and X. Pan, Phys. Med. Biol. **53**, 4777–4799 (2008).

[70] D. Stalling, M. Westerhoff, and H. C. Hege, "amira: A Highly Interactive System for Visual Data Analysis" in C. D. Hansen, C. R. Johnson, eds. *The Visualization Handbook* ch. 38 (Elsevier: Amsterdam) 2005.

[71] N. Otsu, IEEE Trans. SMC **9**, 62-66 (1979).

[72] C. V. Iancu, E. R. Wright, J. B. Heymann, and G. J. Jensen, J. Struct. Biol. **153**, 231-240 (2006).

[73] A. Meents, S. Gutmann, A. Wagner, and C. Schulze-Briese, PNAS **107**, 1094-1099 (2010).

[74] M. L. Taheri, E. A. Stach, I. Arslan, P. A. Crozier, B. C. Kabius, T. LaGrange, A. M. Minor, S. Takeda, M. Tanase, J. B. Wagner, and R. Sharma, Ultramicroscopy **170**, 86-95 (2016).

[75] A. G. Michette, C. J. Buckley, S. J. Pfauntsch, N. R. Arnot, J. Wilkinson, Z. Wang, N. I. Khaleque, and G. S. Dermody, AIP Conf. Proc. **507**, 420-423 (2000).

[76] B. Zhao, J. P. Howard-Knight, A. D. L. Humphris, L. Kailas, E. C. Ratcliffe, S. J. Foster, and J. K. Hobbs, Rev. Sci. Instrum. **80**, 093707 (2009).

[77] M. D. de Jonge, C. G. Ryan, and C. J. Jacobsen, J. Synchrotron Rad. **21**, 1031-1047 (2014).

[78] S. Fakra, A. L. D. Kilcoyne, and T. Tyliszczak, AIP Conf. Proc. **705**, 973-976 (2004).

# Supplemental material for

# Cryo scanning transmission X-ray microscope optimized for spectrotomography

A. F. G. Leontowich,[1] R. Berg,[1] C. N. Regier,[1] D. M. Taylor,[1] J. Wang,[1] D. Beauregard,[1] J. Geilhufe,[1] J. Swirsky,[1] J. Wu,[2] C. Karunakaran,[1] A. P. Hitchcock,[2] and S. G. Urquhart[3]

1. Canadian Light Source Inc., Saskatoon, Saskatchewan S7N 2V3, Canada
2. Dept. of Chemistry & Chemical Biology, McMaster University, Hamilton, Ontario L8S 4M1, Canada
3. Dept. of Chemistry, University of Saskatchewan, Saskatoon, Saskatchewan S7N 5C9, Canada

Corresponding author: adam.leontowich@lightsource.ca

## Supplemental Table I

Selected parameters of commercial stages utilized in the CLS cryo-STXM.

| Component | Direction of motion | Vendor | Stage | Range (mm) | Motor | Encoder | Controller | Driver |
|---|---|---|---|---|---|---|---|---|
| Zone plate | Fine X | PI | P-621.1CD | 0.12 | Piezo - Flexure | Capacitive sensor | PI E-712K162, Ch. 1 | Integrated in controller |
| | Fine Y | PI | P-621.ZCD | 0.14 | Piezo - Flexure | Capacitive sensor | PI E-712K162, Ch. 2 | Integrated in controller |
| | Z | Nanomotion | FB100-60 M4 | 60 | Piezo - Standing wave | Optical increm. | ACS SPiiPlus NTM-4, UDIhp | Nanomotion AB1A-2-HRT-E4 |
| OSA | X | PI | LPS-45 | 13 | Piezo - Slip stick | Optical increm. | PI E-871.1A1 | Integrated in controller |
| | Y | PI | LPS-45 | 13 | Piezo - Slip stick | Optical increm. | PI E-871.1A1 | Integrated in controller |
| | Z | PI | LPS-45 | 26 | Piezo - Slip stick | Optical increm. | PI E-871.1A1 | Integrated in controller |
| Sample | Coarse X | PI | LS-120 | 40 | Phytron VSS-UHV | Optical increm. | Newport XPS-Q8, Ch. 1 | XPS-DRV01 |
| | Coarse Y | Steinmeyer | HT-160 | 16 | Phytron VSS-UHV | Optical increm. | Newport XPS-Q8, Ch. 2 | XPS-DRV01 |
| | Fine X | PI | P-621.1CD | 0.12 | Piezo - Flexure | Capacitive sensor | PI E-712K162, Ch. 3 | Integrated in controller |
| | Fine Y | PI | P-621.ZCD | 0.14 | Piezo - Flexure | Capacitive sensor | PI E-712K162, Ch. 4 | Integrated in controller |
| Detector | X | PI | LS-65 | 52 | Phytron VSS-UHV | Optical increm. | Newport XPS-Q8, Ch. 3 | XPS-DRV01 |
| | Y | Steinmeyer | HT-160 | 16 | Phytron VSS-UHV | Optical increm. | Newport XPS-Q8, Ch. 4 | XPS-DRV01 |
| | Z | PI | LS-65 | 52 | Phytron VSS-UHV | Optical increm. | Newport XPS-Q8, Ch. 5 | XPS-DRV01 |
| Goniometer | X | JEOL | JEM 2010F | 2.25 | Faulhaber AM1524 | Magnetic increm. | MaxNet, Ch. 1 | Oriental Motor RBD-245A-V |
| | Y | JEOL | JEM 2010F | 1.5 | Faulhaber AM1524 | Magnetic increm. | MaxNet, Ch. 2 | Oriental Motor RBD-245A-V |
| | Z | JEOL | JEM 2010F | 1 | Faulhaber AM1524 | Magnetic increm. | MaxNet, Ch. 3 | Oriental Motor RBD-245A-V |
| | θ | JEOL | JEM 2010F | ±80° | Faulhaber AM1524 | Magnetic increm. | MaxNet, Ch. 4 | Oriental Motor RBD-245A-V |

## Supplemental Figure 1

X-ray transmission images at 694.0 eV of a polymer electrolyte membrane hydrogen fuel cell thin section demonstrating different forms of coordinated ZP-OSA scanning. All images were $100 \times 100$ pixels (900 nm pixel size), with 0.9 ms dwell time per pixel, and are all on the same spatial scale. The images were recorded using a 25 nm $\Delta r$, 240 µm $D$ ZP with a 90 µm diameter integrated central stop, and a 70 µm diameter OSA.

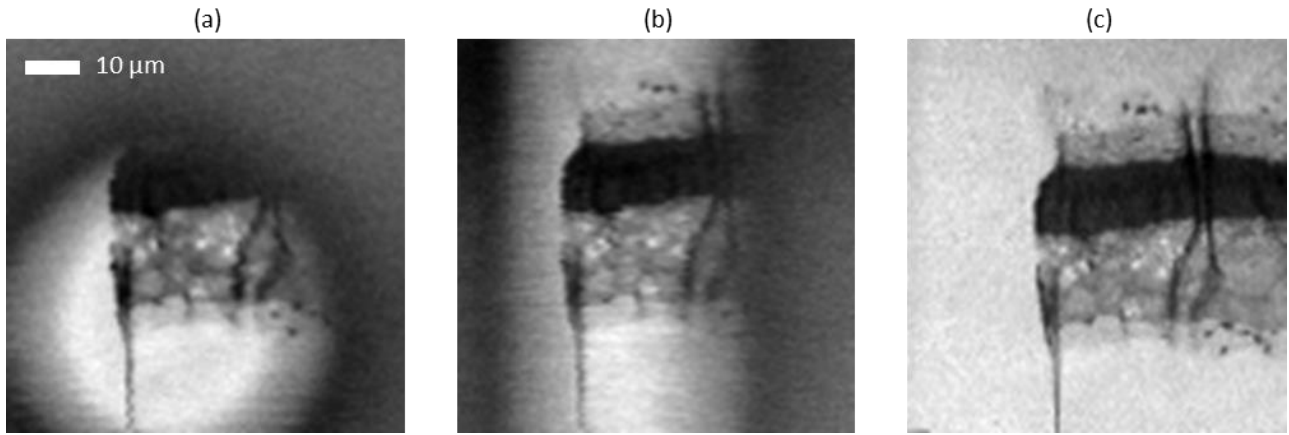

A) ZP fly scanning with a stationary OSA. The diameter of the bright area is 70 µm, which is equal to the diameter of the OSA. This scan mode is generally used when the desired field of view is less than the difference between the diameter of the ZP central stop and the diameter of the OSA, typically less than 20 µm × 20 µm.

B) ZP fly scanning with OSA Y tracking (the OSA moves vertically with each successive horizontal ZP scan line). This scan mode is generally used when the desired field of view is greater than the difference between the diameter of the ZP central stop and the diameter of the OSA, typically greater than 20 µm × 20 µm.

C) ZP point-by-point scanning with OSA X and Y tracking. The OSA moves horizontally and vertically, fully tracking the ZP movement. Although this mode provides the best image quality when ZP scanning fields of view greater than 20 µm × 20 µm, the imaging speed is about an order of magnitude greater than the other two scan types. The OSA stage controllers currently used (Supplemental Table 1) are incompatible with triggering (§2.5) to synchronize fly scanning of the OSA to the ZP. The coordination of point by point scanning the ZP and OSA currently results in significant communication overhead which increases the image acquisition time for this mode.

These examples demonstrate that 90 µm × 90 µm ZP scans are feasible and the ZP illumination does appear to be homogeneous at this photon energy. Further investigation is required to determine whether or not full spatial resolution is preserved all the way out to the edges of such large images, but at this level (900 nm pixel size) there does not appear to be dramatic losses. Often, full 30 nm spatial resolution is not important when users make large images greater than 20 µm × 20 µm. Large images are typically survey images to find a region of interest <20 µm, which is then interrogated at maximum spatial resolution.

**Supplemental Figure 2**

Simplified diagram of the vacuum system.

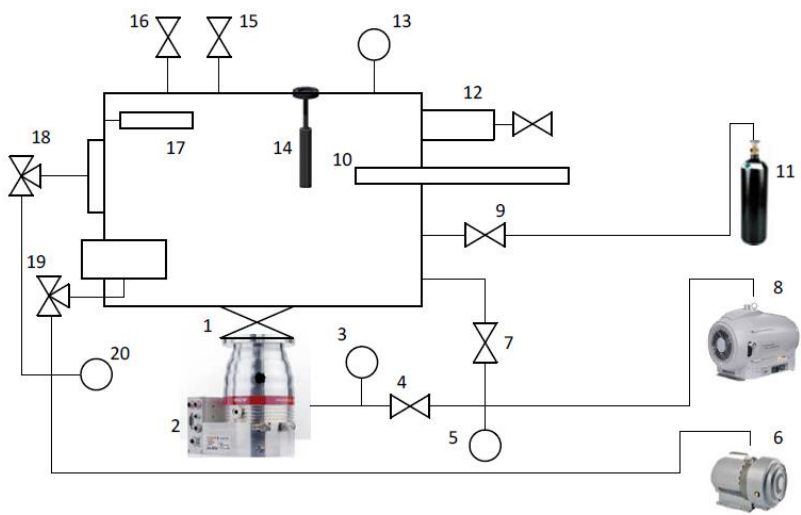

1. Turbo gate valve (VAT 10844-UE08)
2. Turbopump (Pfeiffer HiPace 700M)
3. Turbo foreline gauge (Agilent PVG-500)
4. Turbo foreline valve (Agilent X3202-60032)
5. Rough line gauge (Agilent PVG-500)
6. Load lock rough pump (Agilent SH-110)
7. Roughing valve (MDC 310029)
8. Main rough pump (Agilent IDP-15)
9. Gas vent valve (Swagelok SS-BNS4-DU)
10. Silicon nitride window (SPI 4112SN-BA)
11. Gas cylinder (typically $N_2$)
12. Plasma asher (ibss Group GV10x)
13. Chamber vacuum gauge (Agilent FRG-702)
14. LN2 cold finger (custom)
15. Overpressure check valve (Swagelok SS-4C-1)
16. Burst disk (MDC 420033)
17. Bake out lamps (RBD Instruments IRB100APR)
18. O-ring 3 way valve (Swagelok SS-43GXS4)
19. Load lock 3 way valve (Swagelok SS-43GXS4)
20. Load lock gauge (Agilent PVG-500)

**Supplemental Figure 3**

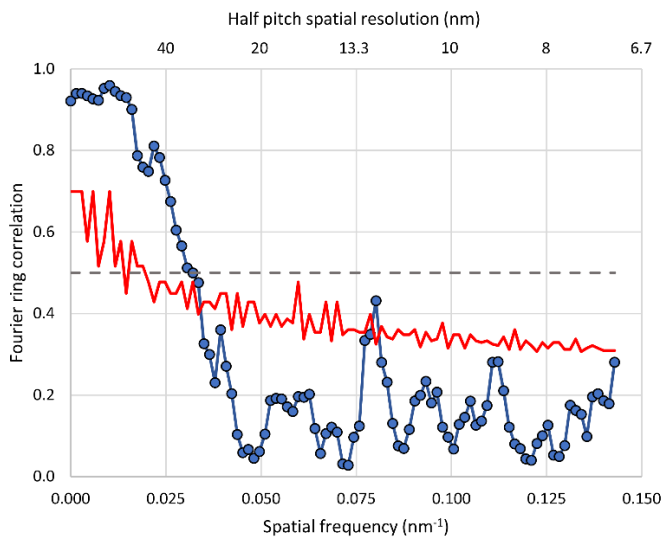

Fourier ring correlation (FRC) analysis of the image presented in Figure 8. The blue data points are the FRC. The dashed grey line represents the 0.5 threshold, and the FRC crosses the 0.5 threshold at a spatial frequency corresponding to a half pitch spatial resolution of 31 nm. The red line is the half-bit threshold, and the FRC crosses the half-bit threshold at a spatial frequency corresponding to a half pitch spatial resolution of 29 nm.

**Supplemental Figure 4**

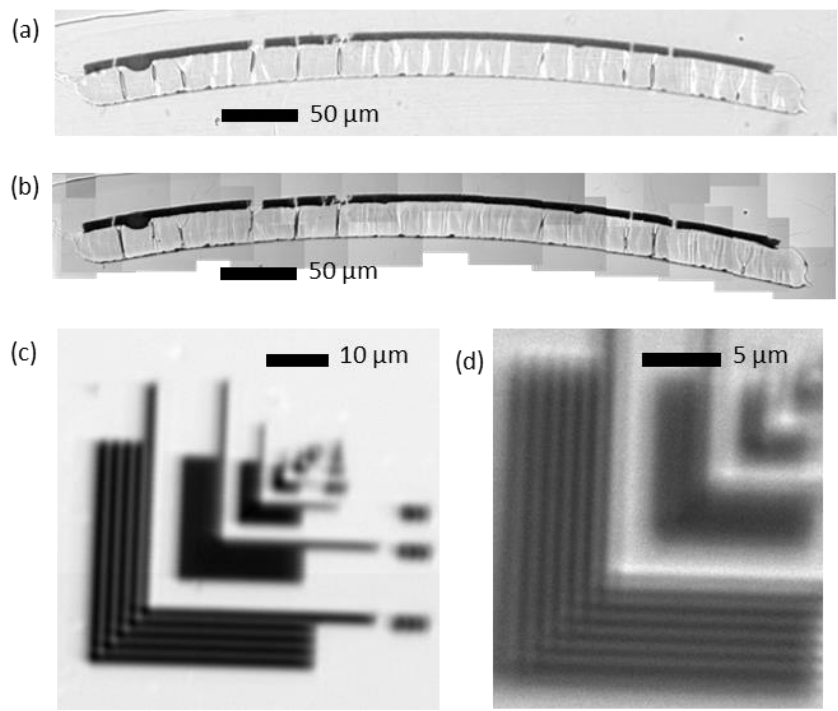

A) Conventional full field optical microscope image (Olympus BX51) of a polymer electrolyte membrane hydrogen fuel cell thin section (100× total magnification, transmission using a white light source). The optically transparent membrane region is on the bottom while the relatively opaque and thinner cathode region is on top.

B) Lens-scanned transmission image of the same sample as A), acquired using the STXM laser light scanning mode. This is a composite of 34 individual images. Visible light from a class IIIA laser diode is directed through a viewport upstream of the STXM and aligned with the X-ray beam axis using two planar mirrors which are in and out of vacuum. The beam passes through the 1.5 mm × 1.5 mm silicon nitride vacuum window and is focused by an optical lens which temporarily replaces the ZP. This image was collected using a 405 nm laser. The lens had a focal length of 4.5 mm, and an effective numerical aperture of 0.16, limited by the vacuum window.

C) Sample-scanned transmission image of a gold on silicon nitride test pattern. This image was collected using a 632 nm laser. The lens had a focal length of 1.5 mm, and an effective numerical aperture of 0.45, limited by the vacuum window. The lines and spaces of a set of elbows with 1 µm half pitch are resolved. The lines and spaces of the next set of elbows with 500 nm half pitch are not resolved.

D) Lens-scanned transmission image of the same sample as C). This image was collected using a 405 nm laser and the system was optimized for this wavelength. The lens had a focal length of 1.5 mm, and an effective numerical aperture of 0.45, limited by the vacuum window. The lines and spaces of a set of elbows with 500 nm half pitch are resolved. The lines and spaces of the next set of elbows with 250 nm half pitch are not resolved.

**Supplemental Figure 5**

All 50 PFSA and carbon support maps used for the spectrotomographic reconstruction in Figure 11. All images are of the same 8 µm × 8 µm field of view.

PFSA maps

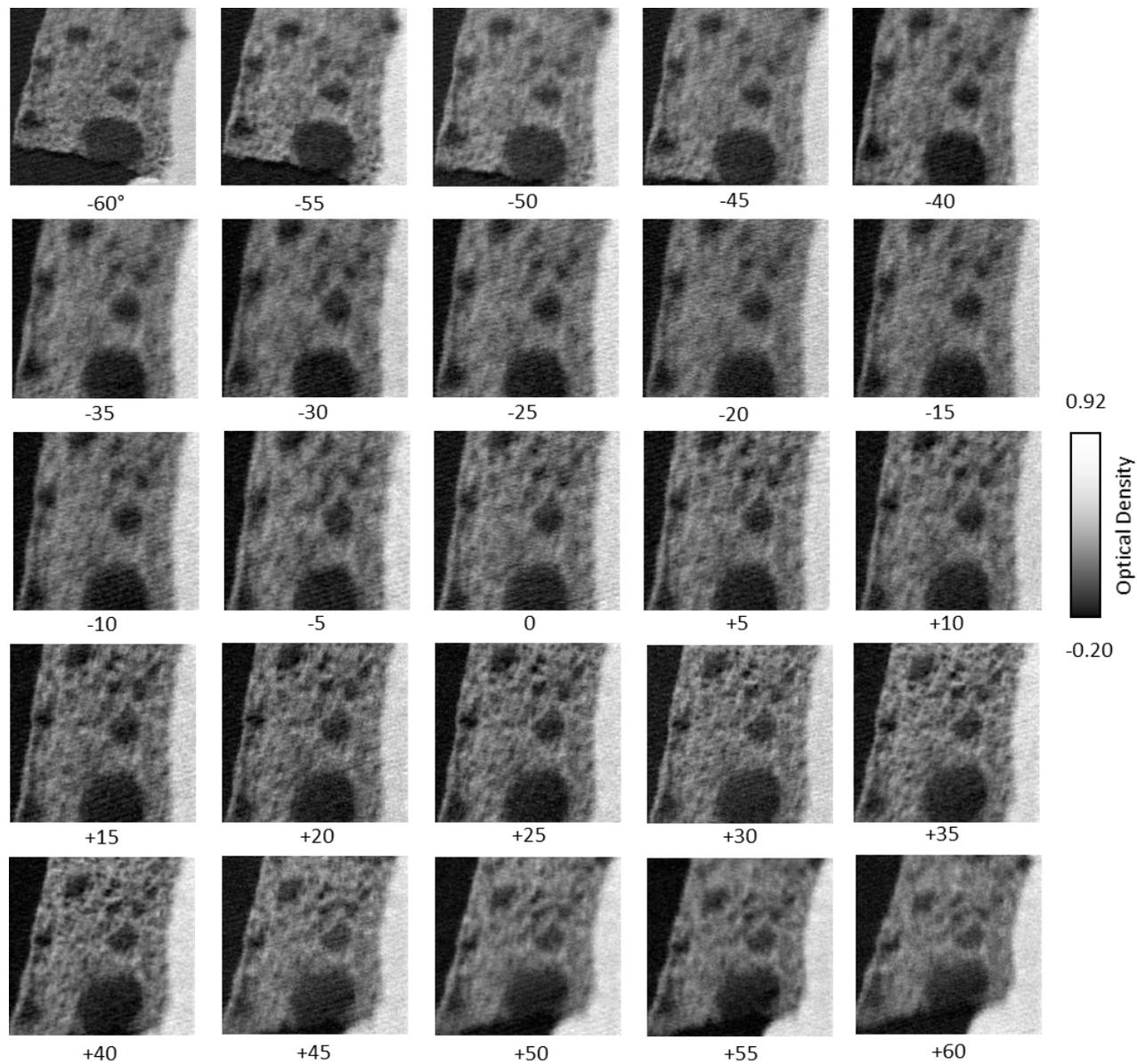

Carbon support maps

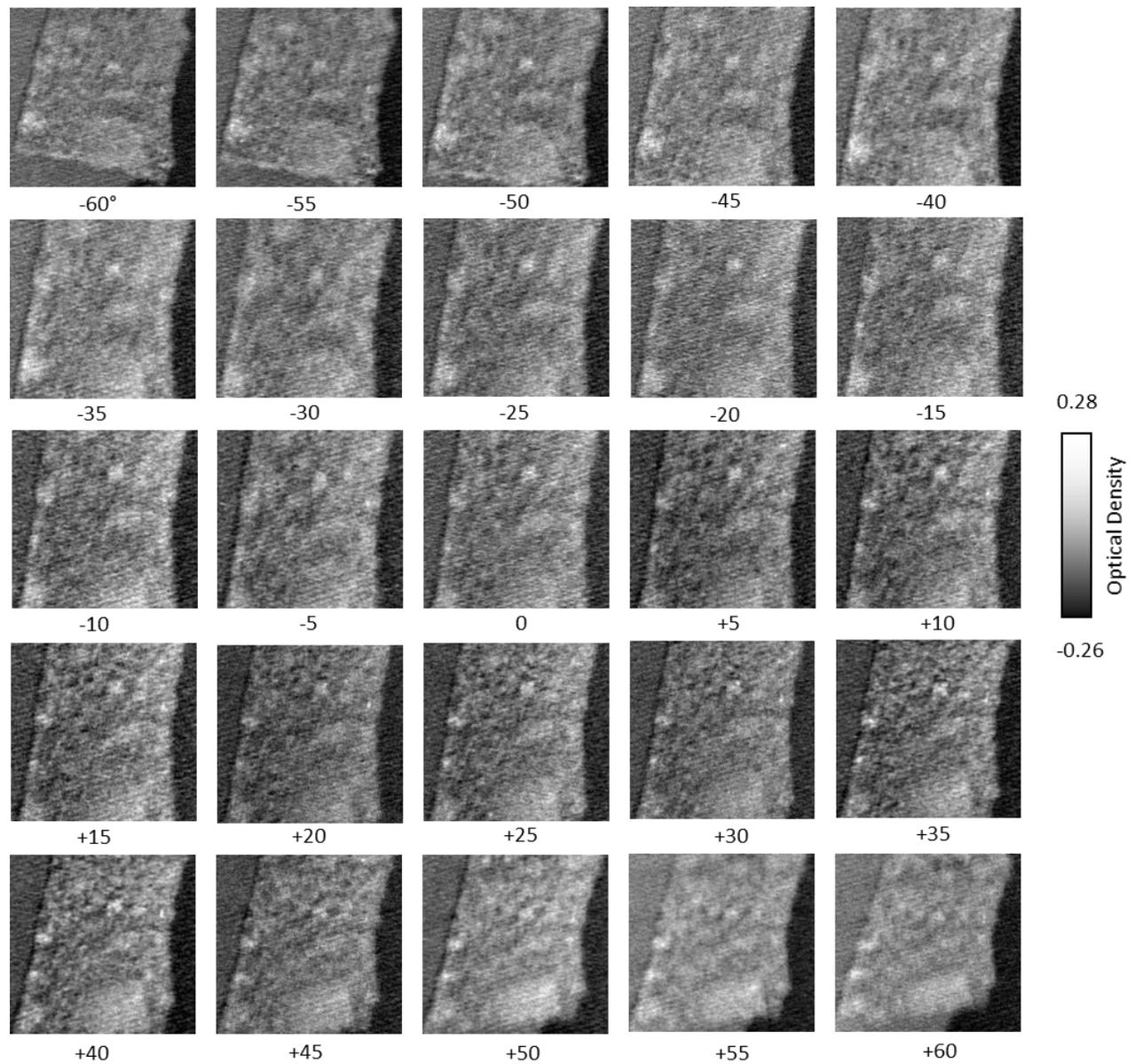

**Supplemental Figure 6**

Slices along the Z thickness direction of the spectrotomogram in Figure 11, with PFSA in green and carbon support in blue.

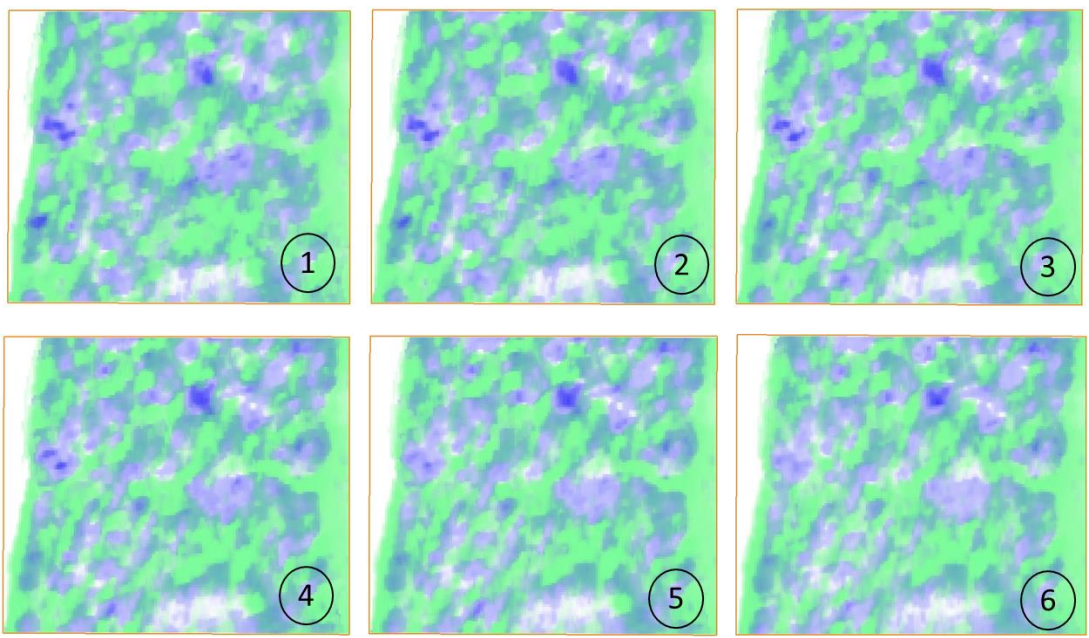